%% file: EXO-17-012_temp.tex
\pdfoutput=1

\documentclass[11pt,twoside,a4paper,cmspaper,final,collab]{cms-tdr}
\def\svnVersion{462447P}\def\cmsCernNoTag{CERN-EP-2018-006}\def\cmsCernDate{\today}\def\cmsMessage{Published in Physical Review Letters as \href{http://dx.doi.org/10.1103/PhysRevLett.120.221801}{\doi{10.1103/PhysRevLett.120.221801.}}}

\begin{document}\cmsNoteHeader{EXO-17-012}

\hyphenation{had-ron-i-za-tion}
\hyphenation{cal-or-i-me-ter}
\hyphenation{de-vices}
\RCS$HeadURL: svn+ssh://svn.cern.ch/reps/tdr2/papers/EXO-17-012/trunk/EXO-17-012.tex $
\RCS$Id: EXO-17-012.tex 460289 2018-05-15 07:49:44Z lshchuts $

\newlength\cmsFigWidth
\newlength\cmsTabSkip\setlength{\cmsTabSkip}{1ex}

\ifthenelse{\boolean{cms@external}}{\providecommand{\NA}{\ensuremath{\cdots}\xspace}}{\providecommand{\NA}{\ensuremath{\text{---}}\xspace}}
\ifthenelse{\boolean{cms@external}}{\setlength\cmsFigWidth{0.85\columnwidth}}{\setlength\cmsFigWidth{0.4\textwidth}}
\ifthenelse{\boolean{cms@external}}{\providecommand{\cmsLeft}{top\xspace}}{\providecommand{\cmsLeft}{left\xspace}}
\ifthenelse{\boolean{cms@external}}{\providecommand{\cmsRight}{bottom\xspace}}{\providecommand{\cmsRight}{right\xspace}}
\ifthenelse{\boolean{cms@external}}{\providecommand{\CL}{C.L.\xspace}}{\providecommand{\CL}{CL\xspace}}
\ifthenelse{\boolean{cms@external}}
{\providecommand{\suppMaterial}{the supplemental material~\cite{supp}}}
{\providecommand{\suppMaterial}{Appendix~\ref{app:suppMat}}}

\newcommand{\MT}{\ensuremath{M_\text{T}}\xspace}
\newcommand{\Mll}{\ensuremath{M_{\ell\ell}}\xspace}
\newcommand{\Mtril}{\ensuremath{M_{3\ell}}\xspace}
\newcommand{\minMOS}{\ensuremath{M_{2\ell\text{OS}}^{\min}}\xspace}
\newcommand{\WZ}{\ensuremath{\PW\PZ}\xspace}
\newcommand{\relIso}{\ensuremath{I_\text{rel}}\xspace}
\newcommand{\N}{\ensuremath{\cmsSymbolFace{N}}\xspace}
\newcommand{\VeN}{\ensuremath{V^{}_{\Pe \N}}\xspace}
\newcommand{\VmN}{\ensuremath{V^{}_{\mu \N}}\xspace}
\newcommand{\VtN}{\ensuremath{V^{}_{\Pgt \N}}\xspace}
\newcommand{\VlN}{\ensuremath{V^{}_{\ell \N}}\xspace}
\newcommand{\mN}{\ensuremath{m_{\N}}\xspace}
\newcommand{\mW}{\ensuremath{m_{\PW}}\xspace}

\cmsNoteHeader{EXO-17-012}

\title{Search for heavy neutral leptons in events with three charged leptons in
proton-proton collisions at \texorpdfstring{$\sqrt{s} =  13\TeV$}{sqrt(s) = 13 TeV}}

\date{\today}

\abstract{
A search for a heavy neutral lepton \N of Majorana nature decaying into a \PW\ boson and a charged lepton is performed
using the CMS detector at the LHC. The targeted signature consists of three prompt charged leptons in any flavor
combination of electrons and muons. The data were collected in proton-proton collisions at a center-of-mass
energy of 13 TeV, with an integrated luminosity of 35.9\fbinv. The search is performed in the \N mass range
between 1\GeV and 1.2\TeV. The data are found to be consistent with the expected standard model background.
Upper limits are set on the values of $\abs{\VeN}^2$ and $\abs{\VmN}^2$, where $\VlN$
is the matrix element describing the mixing of \N with the standard model neutrino of flavor $\ell$.
These are the first direct limits for \N masses above 500\GeV and the first limits obtained at a hadron collider
for \N masses below 40\GeV.
}

\hypersetup{%
pdfauthor={CMS Collaboration},%
pdftitle={Search for heavy neutral leptons in events with three charged leptons in  proton-proton collisions at 13 TeV},%
pdfsubject={CMS},%
pdfkeywords={CMS, physics, HNL}}

\maketitle

The standard model (SM) of particle physics has been successful in describing many phenomena, however several
observations remain unexplained, including the nature of dark matter (DM), the baryon asymmetry of the universe,
and the smallness of neutrino masses. The latter can be naturally accommodated by the so-called ``seesaw"
mechanism~\cite{seesawI_1, seesawI_2, seesawI_3, seesawI_4, seesawII_1,seesawII_2,seesawII_3,seesawII_4,seesawII_5,seesawII_6, seesawIII},
in which a new heavy neutral Majorana lepton (or heavy neutrino) \N is postulated.

A model that incorporates the seesaw mechanism, while also providing a DM candidate and giving
a possible explanation for the baryon asymmetry, is known as the neutrino minimal standard model ($\nu$MSM).
In this model, three right-handed heavy sterile neutrinos are added to 
the SM~\cite{Appelquist:2002me,Appelquist:2003uu,Asaka:2005an,Asaka:2005pn}.
The lightest neutrino, $\N_1$,  can explain the DM in the universe, while the heavier neutrinos, $\N_2$ and $\N_3$,
can be responsible for the baryon asymmetry through leptogenesis~\cite{vMSM_lg,vMSM_no,Asaka:2005pn,Canetti:2012kh,Canetti:2014dka,Antusch:2017pkq} 
or neutrino oscillations~\cite{Asaka:2005an,Asaka:2005pn}.

In this Letter, a search for a heavy neutrino decaying into a charged lepton and a $\PW$ boson (either an on-shell $\PW$
or off-shell $\PW^*$) is performed using the CMS detector at the CERN LHC~\cite{Evans:2008zzb}. The heavy neutrinos
are produced through mixing with the SM neutrinos, governed by the parameters $\VeN, \VmN$, and $\VtN$,
where $\VlN$ is the matrix element describing the mixing of \N with the SM neutrino of flavor $\ell$.
The production cross section, decay width, and lifetime of \N depend on $\abs{\VlN}^2$ and its mass $\mN$.
In the $\nu$MSM,  $\N_1$ is expected to be too light and long-lived to produce an unambiguous
signal in the CMS detector, but $\N_2$ and $\N_3$ could decay to $\PW\ell$, $\PZ\nu$, or $\PH\nu$,
and are therefore potentially detectable.

Earlier searches for heavy Majorana neutrinos at the LHC have been undertaken by the ATLAS and CMS Collaborations
at $\sqrt{s} = 7$ and $8\TeV$~\cite{Aad:2011vj,ATLAS:2012ak,Chatrchyan:2012fla,CMS:2012zv,Khachatryan:2014dka,Khachatryan:2015gha,Aad:2015xaa,Khachatryan:2016olu},
employing a signature of same-sign dileptons and jets, exploring the mass range  $40 < \mN < 500\GeV$.
Other experiments have searched in the mass region
$\mN < 40\GeV$~\cite{CooperSarkar:1985nh,Bergsma:1985is,Badier:1985wg,Bernardi:1987ek,L3,Baranov:1992vq,Vilain:1994vg,Gallas:1994xp,DELPHI,Vaitaitis:1999wq,Acciarri:1999qj,Achard:2001qv,Liventsev:2013zz,Aaij:2014aba}
and precision electroweak measurements provide limits on the mixing parameters independent
of $\mN$~\cite{delAguila:2008pw,Akhmedov:2013hec,deBlas:2013gla,Basso:2013jka,Antusch:2015mia}.
A recent review of constraints can be found in Ref.~\cite{Deppisch:2015qwa}.

This analysis targets \N production in leptonic $\PW^{(*)}$ boson decays, $\PW^{(*)} \to \PN \ell$ ($\ell = \Pe, \mu$),
with subsequent prompt decays of \N to $\PW^{(*)}\ell$, where the vector boson decays to
$\ell\nu$~\cite{delAguila:2008cj,Das:2012ze,Das:2014jxa,Izaguirre:2015pga,Dib:2015oka,Das:2016hof,Dib:2016wge,Dib:2017iva,Dib:2017vux,Dube:2017jgo,Das:2017gke,Arbelaez:2017zqq,Bhardwaj:2018lma}.
The event signature consists of three charged leptons in any combination of electrons and muons,
excluding those events containing three leptons of the same charge.
Because of the presence of a SM $\nu$ escaping detection, a mass peak of \N cannot be reconstructed.
Therefore the search exploits kinematic properties of the three leptons to discriminate between
the signal and SM backgrounds.
These backgrounds consist of events containing leptons from hadron decays, leptons from conversions, and
SM sources of multiple leptons such as diboson production or top quark (pair) production in association with a boson.
Exploiting the trilepton topology allows the mass range $1\GeV < \mN < 1.2\TeV$ to be explored
using $\Pp\Pp$ collision data collected by the CMS experiment at $\sqrt{s}=13\TeV$,
corresponding to an integrated luminosity of 35.9\fbinv.

The central feature of the CMS apparatus~\cite{Chatrchyan:2008zzk} is a superconducting solenoid of 6\unit{m} diameter,
providing a magnetic field of 3.8\unit{T}. A silicon pixel
and strip tracker, a lead tungstate crystal electromagnetic calorimeter (ECAL), and a brass and scintillator
hadron calorimeter, each composed of a barrel and two endcap sections, reside within the solenoid. Forward calorimeters
extend the pseudorapidity ($\eta$) coverage.
Muons are detected in gas-ionization detectors embedded in the steel flux-return yoke outside the solenoid.
Events of interest are recorded with several triggers~\cite{Khachatryan:2016bia},
requiring the presence of one, two, or three light leptons ($\Pe$ or $\Pgm$), leading to very high efficiency,
nearing 100\% in most kinematic regions of the search.

Samples of simulated events are used to estimate the background from some of the SM processes
and to determine the heavy neutrino signal acceptance.
The SM background samples are produced using the Monte Carlo (MC)
\MGvATNLO 2.2.2 or 2.3.3 generator~\cite{MADGRAPH5}
at leading order (LO) or next-to-leading order (NLO) in perturbative quantum chromodynamics,
with the exception of $\Pg\Pg\to\PZ\PZ$, which is simulated at LO with \MCFM~7.0~\cite{Campbell:2010ff},
and all other diboson production processes, which are generated at NLO with
the \POWHEG~v2~\cite{Melia:2011tj,Nason:2013ydw} generator.

The NNPDF3.0~\cite{Ball:2014uwa} LO (NLO) parton distribution functions (PDFs) are used for the simulated samples generated
at LO (NLO). Parton showering and hadronization are described using the \PYTHIA 8.212 generator~\cite{Sjostrand:2007gs}
with the CUETP8M1 underlying event tune~\cite{Skands:2014pea,CMS-PAS-GEN-14-001}. Double counting of the partons generated
with \MGvATNLO and \PYTHIA is removed using the MLM~\cite{Alwall:2007fs} and \textsc{FxFx}~\cite{Frederix:2012ps}
matching schemes in the LO and NLO samples, respectively.

Signal samples are generated with \MGvATNLO\!2.4.2 at NLO precision, following the implementation
of Ref.~\cite{Degrande:2016aje}. They include processes leading to
\N production via the charged-current Drell--Yan (DY) process, gluon fusion,
and $\PW\gamma$ fusion. The latter production mechanism is important
for  $\mN > 600\GeV$~\cite{Alva:2014gxa}. The first two production mechanisms employ
the NNPDF3.0 NLO PDF set~\cite{Ball:2014uwa}, while the last uses the
LUXqed\_plus\_PDF4LHC15\_nnlo\_100 PDF set~\cite{Manohar:2016nzj}.
Parton showering and hadronization are simulated with \PYTHIA.
Only the final states with three leptons (electrons or muons) and a neutrino
are generated.

The effects of additional $\Pp\Pp$ interactions in the same or adjacent $\Pp\Pp$ bunch crossings (pileup) are accounted
for in the simulations. The MC generated events include the full simulation of the CMS detector based
on \GEANTfour~\cite{Geant} and are reconstructed using the same CMS software as used for data.

Information from all subdetectors is combined offline by the CMS particle-flow algorithm~\cite{Sirunyan:2017ulk}
used to reconstruct and identify individual particles and to provide a global description of the event.
The particles are classified into charged hadrons, neutral hadrons, photons, electrons, and muons.

Jets are reconstructed using the anti-$\kt$ clustering algorithm~\cite{Cacciari:2008gp} with a distance parameter of 0.4,
as implemented in the \FASTJET package~\cite{Cacciari:fastjet1,Cacciari:fastjet2}.
Jet energies are corrected for residual nonuniformity and nonlinearity of the detector response
using simulated and collision data event samples~\cite{Chatrchyan:2011ds,Khachatryan:2016kdb,CMS-PAS-JME-16-003}.

To identify jets originating from b quarks, the combined secondary vertex CSVv2 
algorithm~\cite{Chatrchyan:2012jua,BTV-16-002} is used.
This has an efficiency of approximately 80\% for tagging a b quark jet, 
and a mistagging rate of 10\% for light-quark and gluon jets, and about 40\% for c quark jets.
Jets with $\pt > 25\GeV$ and $\abs{\eta}<2.4$ are considered $\cPqb$ quark jets (``$\cPqb$ jets")
if they satisfy the loose working point requirements~\cite{BTV-16-002} of this algorithm.
Events with one or more identified $\cPqb$ jets are vetoed in the analysis to reduce the \ttbar background.

The missing transverse momentum $\ptmiss$ is defined as the magnitude of the negative vector sum
$\ptvecmiss$ of the transverse momenta of all reconstructed particles in the event, taking into account 
jet energy corrections~\cite{CMS-PAS-JME-16-004}.

The primary $\Pp\Pp$ interaction vertex (PV) is taken to be the reconstructed vertex with
the largest value of summed physics-object $\pt^2$.
The physics objects are the jets, clustered using the jet finding algorithm~\cite{Cacciari:2008gp,Cacciari:fastjet1}
with the tracks assigned to the vertex as inputs, and the associated missing transverse momentum,
taken as the negative vector sum of the \pt of those jets.

The analysis depends crucially on identifying electrons and muons, with good efficiency and low contamination.
Electrons are reconstructed by combining the information from the ECAL  and the tracker~\cite{Khachatryan:2015hwa}.
Electrons are required to be within the tracking system volume, $\abs{\eta} < 2.5$, and have a minimum \pt of 10\GeV.
Electron identification is performed using a multivariate 
discriminant that includes the shower shape and track quality information.
To reject electrons originating from photon conversions
in detector material, electrons must have measurements in all innermost layers of the tracking system
and must not be matched to any secondary vertex~\cite{Khachatryan:2015hwa}.

Muons are reconstructed by combining the information from the tracker and
muon spectrometer in a global fit~\cite{Chatrchyan:2012xi}.
The quality of the geometrical matching between measurements made separately in the two systems
is used in the further selection of muons.
Only muons within the muon system acceptance of $\abs{\eta}<2.4$ and with $\pt > 5\GeV$ are considered.
Electrons within a cone of $\Delta R = \sqrt{\smash[b]{(\Delta\eta)^2+(\Delta\phi)^2}} <0.05$
of a muon are discarded as those likely coming from radiation.

To ensure that electron and muon candidates are consistent with originating from the PV,
the transverse (longitudinal) impact parameter of the leptons with respect to this vertex
must not exceed 0.5 (1.0)\unit{mm}, and the displacement divided by its uncertainty must not exceed 4.

Leptons originating from decays of heavy particles, such as electroweak bosons or \N, are referred to as ``prompt'',
while leptons produced in hadron decays are called ``nonprompt''.
For convenience, we also include misidentified hadrons and jets in the nonprompt-lepton classification.
A powerful discriminator between these two types of leptons is the isolation variable \relIso.
It is defined as the pileup-corrected scalar $\pt$ sum of particles within a cone of $\Delta R<0.3$ around
the lepton candidate's direction at the vertex, divided by the lepton candidate \pt.
The summation comprises the reconstructed charged hadrons originating from the PV, neutral hadrons, and photons.

Electrons and muons that pass all the aforementioned requirements and satisfy $\relIso < 0.6$ are referred to as ``loose leptons''.
Leptons that additionally satisfy $\relIso < 0.1$ and, in the case of electrons, pass a more stringent requirement on
the multivariate discriminant, chosen to maximize the signal over background ratio, are referred to as ``tight leptons''.
Events containing exactly three loose leptons, not all having the same charge, are retained in the analysis.

To distinguish between SM background and \N production, the three leptons are required
to pass the tight selection, and the following variables are used:
the flavor, charge, and \pt of the leptons in the event;
the invariant mass of the trilepton system $\Mtril$;
the minimum invariant mass of all opposite-sign lepton pairs in the event $\minMOS$;
and the transverse mass $\MT = \sqrt{\smash[b]{2\pt^\ell\ptmiss[1-\cos(\Delta\phi)]}}$,
where $\pt^\ell$ is the transverse momentum of the lepton that is not used in the $\minMOS$ calculation,
and $\Delta\phi$ is the azimuthal angle between $\vec{p}_\text{T}^\ell$ and $\ptvecmiss$.

To address the kinematically distinct cases of \N masses below and above that of the $\PW$ boson,
two search regions are defined, referred to as the low- and high-mass regions.

In the low-mass region (targeting $\mN < m_\PW$), \N is produced in the decay of an on-shell $\PW$ boson,
leading to the decay $\PW \to \ell\ell'\ell''\nu$ via an intermediate \N.
To reflect the targeted $\mN$ range in this region, and to suppress
the background from $\PZ\to \ell^+ \ell^-$ production with an accompanying high-\pt lepton from an asymmetric photon conversion,
the requirement $\Mtril < 80\GeV$ is imposed. The background from
$\PW \gamma^*$ events, with $\gamma^* \to \ell^+ \ell^-$, is reduced by rejecting events 
that contain an opposite-sign same-flavor (OSSF) lepton pair.
The effectiveness of this requirement relies on the fact that \N is a Majorana particle
and can decay to a lepton of equal or opposite charge to that of its parent $\PW$ boson.

Events in the low-mass region are required to have $\ptmiss < 75\GeV$ to suppress $\ttbar$ background.
The highest \pt (leading) lepton must satisfy $\pt > 15\GeV$, while the next-to-highest \pt (subleading) lepton
must have $\pt > 10\GeV$.
The third (trailing) lepton must have $\pt > 10\,(5)\GeV$ if it is an electron (muon).
In addition the following conditions are imposed to avoid trigger threshold effects:
in $\Pe\Pgm\Pgm$ events, if a trailing electron has $10<\pt<15\GeV$, the leading muon is required to have $\pt > 23\GeV$, and
if a trailing muon has $5 < \pt < 8 \GeV$, the leading and subleading electrons must satisfy $\pt > 25$ and $15\GeV$;
in $\Pe\Pe\Pgm$ events, if a trailing muon has $\pt > 8\GeV$, either the leading electron must have $\pt > 23\GeV$,
or the subleading electron must have $\pt > 15\GeV$.
These requirements lead to a signal selection efficiency of 5--7\% for a trilepton final state depending on the \N mass.

The events are subdivided into two ranges of leading lepton $\pt$: $15 < \pt^\text{leading} < 30\GeV$ and
$30 < \pt^\text{leading} < 55\GeV$.  The lower range has higher signal efficiencies for $\mN$
close to $\mW$, leading to three leptons with similar $\pt$ spectra.
The higher range targets very low $\mN$ down to $1\GeV$, with one energetic
and two soft leptons in the event.

\begin{figure*}[h!]
\centering
\includegraphics[width=1.0\textwidth]{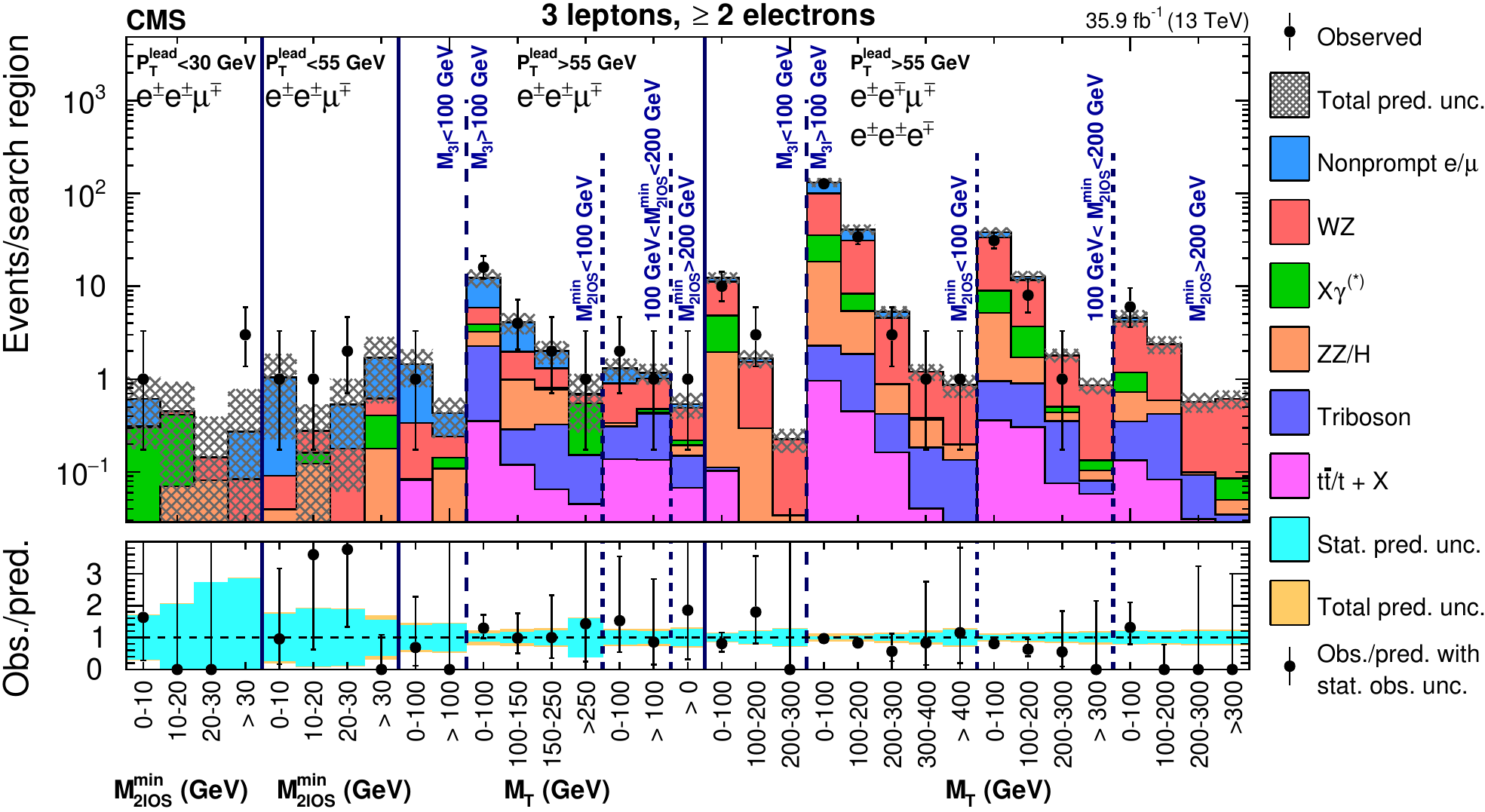} \\
\vspace{0.4cm}
\includegraphics[width=1.0\textwidth]{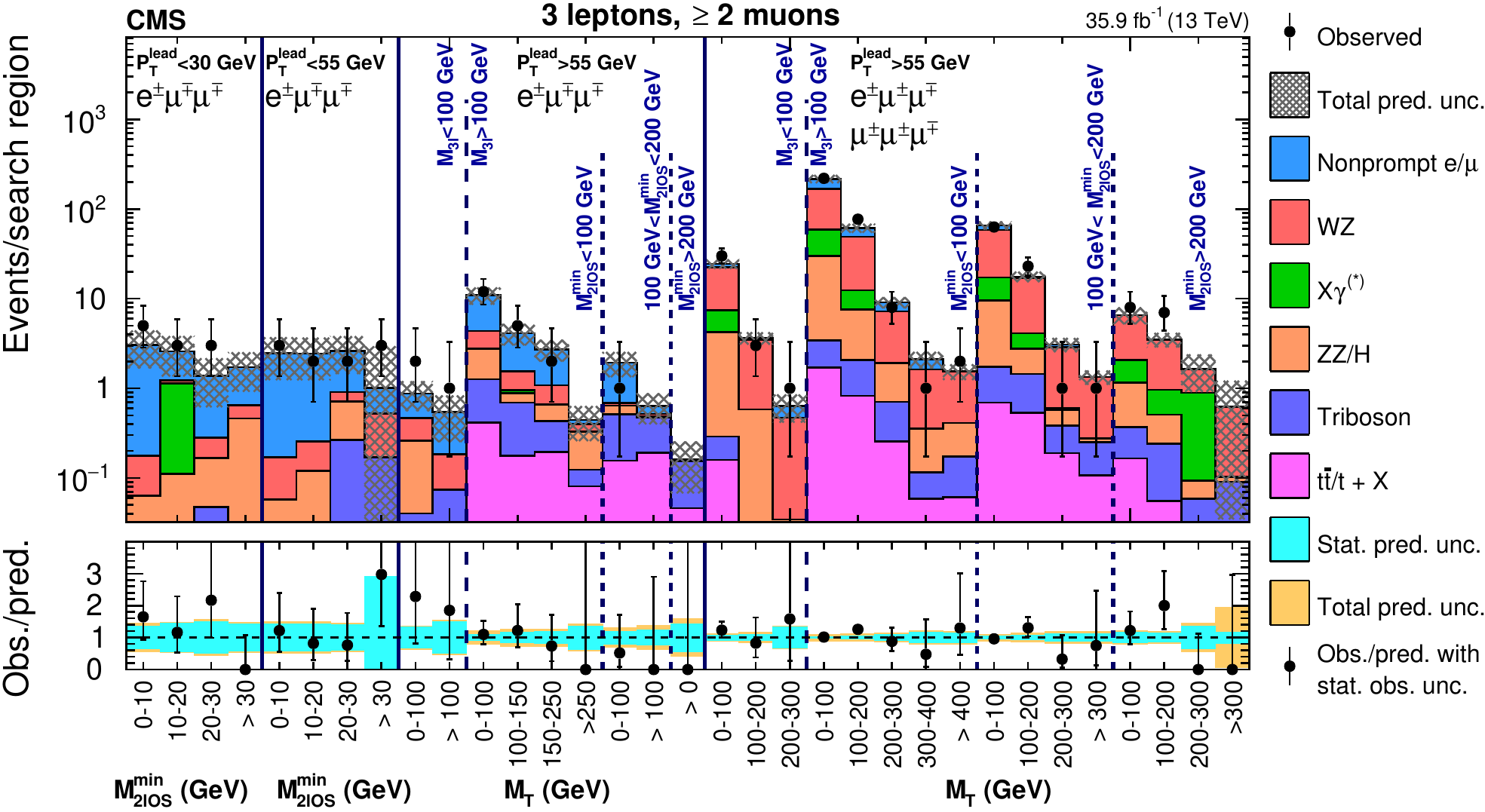}
\caption{Observed and expected event yields as a function of $\minMOS$ and $\MT$ for events with at least two electrons (upper),
and with at least two muons (lower). The contribution of each background source is shown. The first 8 bins of each figure correspond to the low-mass region, while the rest display the high-mass region.
}
\label{fig:SRplots}
\end{figure*}

Finally, the variable $\minMOS$, which is correlated with $\mN$, is used to further subdivide
the events into four bins ($<$10, 10--20, 20--30, and $>$30\GeV) giving a total of eight search regions,
as shown in Fig.~\ref{fig:SRplots}.

In the high-mass region (targeting $\mN > m_\PW$), \N is produced in the decay of a high-mass off-shell $\PW$ boson,
leading to three relatively energetic leptons, and more sizable $\ptmiss$. In this region, the three selected leptons must satisfy
$\pt > 55, 15, 10\GeV$. With these requirements, the background from
$\PW \gamma^*$ production is negligible, and events containing an OSSF lepton pair are therefore retained,
but with the invariant mass of any OSSF lepton pair required to satisfy $\Mll>5\GeV$.
The backgrounds from $\PW\PZ$ and $\PZ\gamma^{(*)}$ production are respectively suppressed
by vetoing events having an OSSF lepton pair with $\abs{\Mll (\text{or} \; \Mtril) -M_\PZ} < 15\GeV$.
Signal selection efficiency for trilepton final state reaches up to 50\%.

In order to improve the discrimination of signal from background, the high-mass region is divided into two event categories:
events containing an OSSF lepton pair and events with no such pair.
Both categories are divided into bins of $\Mtril$ and $\minMOS$, 
each further subdivided according to $\MT$, which tends to be large for high \N masses.
This results in a total of 25 search regions, as shown in Fig.~\ref{fig:SRplots}.

To extract values of $\abs{\VeN}^2$ and $\abs{\VmN}^2$ separately,
the results for both the low- and high-mass regions are obtained for events with
$\ge$2 electrons or $\ge$2 muons, respectively.

We now consider the most important sources of background and their associated systematic uncertainties.
The \ttbar and DY processes, with an additional nonprompt lepton,
constitute the main background for events in the low- and high-mass regions with no OSSF lepton pair.
It is estimated by using the tight-to-loose ratio method described in Ref.~\cite{SUS-15-008}.
The probability for a loose nonprompt lepton to pass the tight selection criteria
is measured in a multijet sample in data enriched in nonprompt leptons.
This probability is applied to events that pass the full signal selection,
but contain at least one lepton that fails the tight selection, while
satisfying the loose selection requirements.
The method is validated using simulation and data in control regions enriched in $\ttbar$ or DY+jets events.
Agreement between the predicted and observed yields in the various control regions is found to
be within the overall systematic uncertainty of 30\% assigned to this background estimate.

The background from $\WZ$ and $\PW \gamma^*$ production
is dominant in the high-mass region containing an OSSF lepton pair.
It is obtained from simulation, with the simulated yield normalized to data in a control region
formed by selecting three tight leptons with $\pt > 25, 15, 10\GeV$,
and requiring an OSSF lepton pair with invariant mass $\Mll$ consistent with a $\PZ$ boson:
$\abs{\Mll-M_\PZ} < 15\GeV$.
In addition, events are required to have $\ptmiss>50\GeV$. The ratio of the predicted to observed $\WZ$ yield in
this control region is found to be $1.08 \pm 0.09$.
This ratio is used to normalize the MC generated event samples,
and its associated uncertainty is propagated to the result.

Production of $\PZ\PZ$ events with both $\PZ$ bosons decaying leptonically,
and one lepton not identified, results in a trilepton signature.
This contribution is estimated from simulation, and the simulated yield is normalized
in a control region containing four leptons that form two OSSF lepton pairs with invariant masses consistent
with a $\PZ$ boson. The ratio of data to simulation in the control region is found to be $1.03\pm0.10$.
An additional uncertainty of 25\% is assigned to the prediction of events with $\MT > 75\GeV$,
based on a comparison of the observed and predicted event yields in the control region.

External and internal photon conversions (X$\gamma^{(*)}$) contribute to the trilepton
final state when a photon is produced with a $\PZ$ boson,
and this photon undergoes an asymmetric conversion in which one of the leptons has very low $\pt$
and fails the lepton selection criteria.
This contribution is obtained from simulation and verified in a data control region enriched
in conversions from the $\PZ$+jets process, with $\PZ \to\ell\ell\gamma^{(*)}$ and
$\gamma^{(*)}\to\ell\ell$, where one of the leptons is outside the detector acceptance.
The control region is defined by $\abs{\Mll-M_\PZ} > 15\GeV$ and $\abs{M_{3\ell} - M_\PZ} < 15\GeV$.
The ratio of data to expected background in the control region
is $0.95\pm0.08$, and is used to normalize the simulation. Kinematic properties of the events
in data are used to set a systematic uncertainty in the photon conversion background of 15\%.

Other SM processes that can yield three or more prompt leptons include triboson production
($\PW$, $\PZ$, $\PH$, or a prompt $\gamma$) and single-boson production in association
with a single top quark or a \ttbar pair (\ttbar/\cPqt+X). Such processes generally have very small
production rates and in some cases are
further suppressed by the $\cPqb$ jet veto. They are estimated from simulation
with an uncertainty of 50\%, which includes uncertainties due to experimental effects,
event simulation, and theoretical calculations of the cross sections.

The background from the mismeasurement of charge arises from events
with an $\Pep\Pem$ pair in which the charge of one of the electrons is misreconstructed.
It is predicted using simulation, which is validated to agree within 10\% of
an estimate obtained from data~\cite{Sirunyan:2017uyt}.

Systematic uncertainties affecting any process whose yield is
estimated from simulation are considered, such as those from trigger efficiency,
lepton selection efficiencies, jet energy scale, $\cPqb$ jet veto efficiency, pileup modeling,
and those related to fixed-order calculations in event simulation. The effect of each uncertainty
on the event yields is computed and accounted for.

The uncertainty in the trigger efficiency is obtained by measuring the efficiencies of all trigger components
using the tag-and-probe technique~\cite{Chatrchyan:2012xi,Khachatryan:2015hwa}, and is estimated to
be 2\% for events with leading lepton $\pt > 30\GeV$ and 5\% otherwise.
Lepton identification efficiencies are also computed using the tag-and-probe technique with an uncertainty of 2\% per lepton.

The impact of the jet energy scale uncertainty is determined by shifting the jet energy correction factors up and down
by their estimated uncertainty for each jet, and recalculating all kinematic quantities obtained from jets.
This results in an uncertainty in event yields of up to 3\%, depending on the search region. Correlation effects due
to the migration of events from one search region to another are also taken into account.
Similarly, the $\cPqb$ jet veto efficiency is corrected for differences between data and simulation,
leading to an uncertainty in event yields of 1--5\%. The uncertainty in yields due to modeling of pileup
is computed by modifying the total inelastic scattering cross section by 5\%~\cite{Sirunyan:2017vio}, and is measured to be 1--5\%,
depending on the search region. The uncertainty in the integrated luminosity is 2.5\%~\cite{CMS-PAS-LUM-17-001}.

Further uncertainties in background yields estimated from simulation arise from the unknown higher-order effects
in the theoretical calculations of cross sections, and from uncertainties in the knowledge of the proton PDFs.
Uncertainties in  the renormalization and factorization scales affect the signal cross section
and acceptance. These are evaluated by independently varying the aforementioned scales up and down
by a factor of two relative to their nominal values.
The uncertainties associated with the choice of PDFs are estimated by considering replica PDF sets generated 
using weights, giving a PDF probability distribution centered on the nominal PDF set~\cite{Butterworth:2015oua}.

The limited statistical precision of the available MC samples leads to an additional uncertainty of 1--30\%, depending on
the process and search region.

The expected and observed yields together with the relative contributions of the different background sources in each search region,
are shown in Fig.~\ref{fig:SRplots}. Tabulated results and enlarged versions of Fig.~\ref{fig:SRplots}, with potential signals
superimposed, are provided in \suppMaterial.
We see no evidence for a significant excess in data beyond the expected SM background.
We compute 95\% confidence level (\CL) upper limits on $\abs{\VeN}^2$ and $\abs{\VmN}^2$ separately, while assuming
other matrix elements to be 0, using the CL$_\text{s}$ criterion~\cite{Junk:1999kv,Read:2002hq} under the asymptotic
approximation for the test statistic~\cite{Cowan:2010js,ATLAS:1379837}.
A simultaneous fit of all search regions is performed and all systematic uncertainties are treated as log-normal nuisance parameters in the fit.

\begin{figure*}[htbp]
\centering
  \includegraphics[width=.49\textwidth]{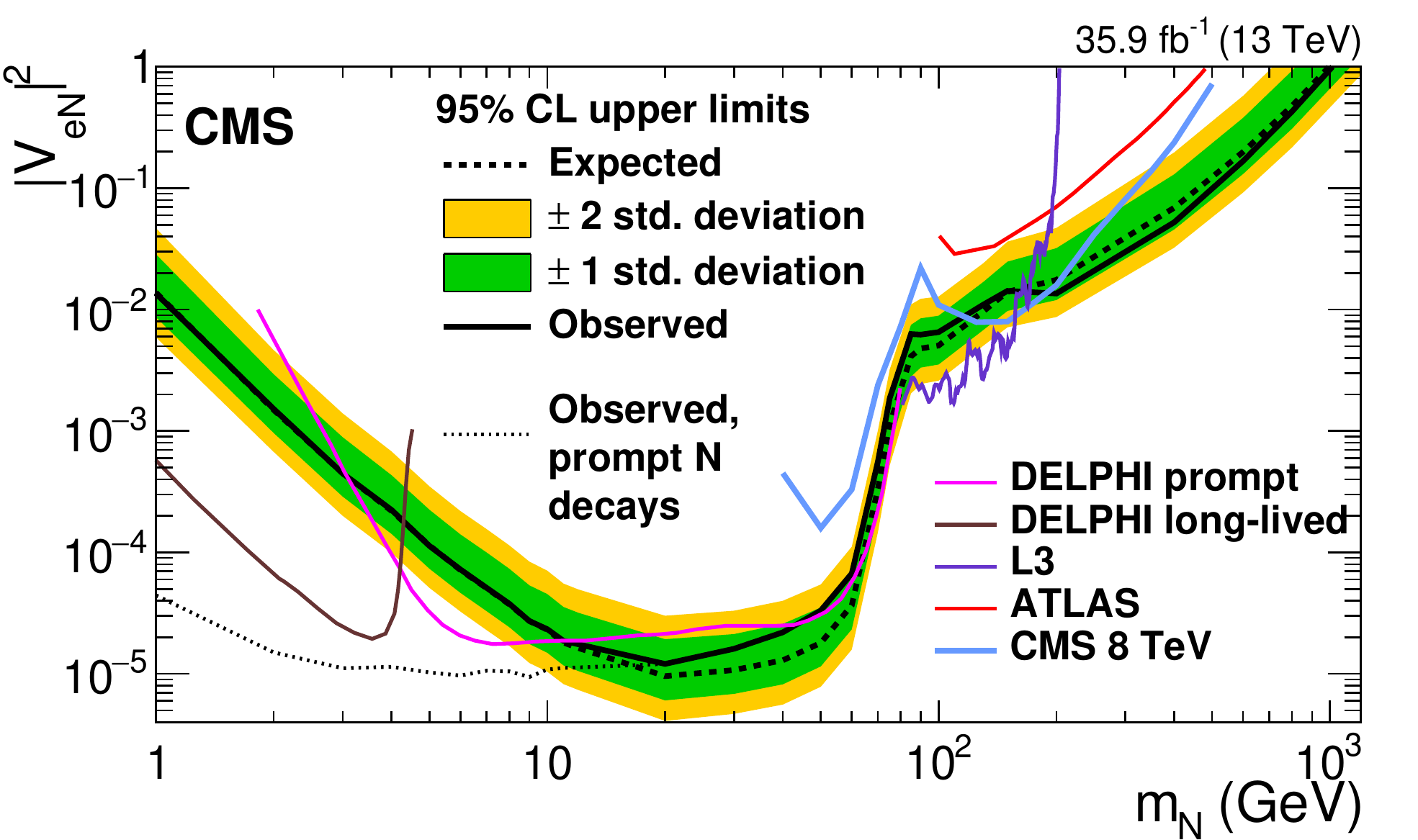}
  \includegraphics[width=.49\textwidth]{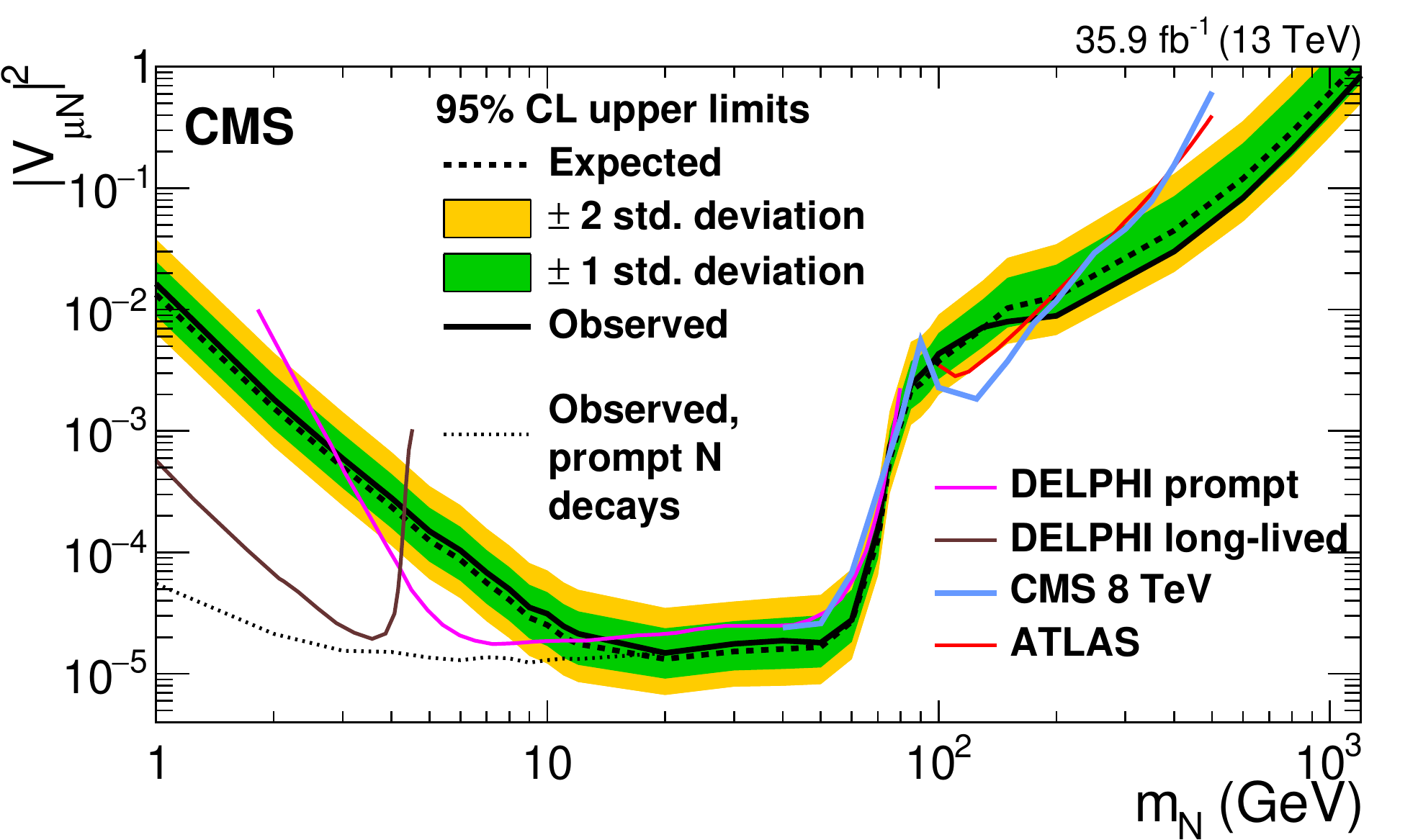}
\caption{Exclusion region at 95\% \CL in the $\abs{\VeN}^2$ vs. $\mN$ (left) and
  $\abs{\VmN}^2$ vs. $\mN$ (right) planes. The dashed black curve is the expected upper limit, with one and two
  standard-deviation bands shown in dark green and light yellow, respectively. The solid black curve is the
  observed upper limit, while the dotted black curve is the observed limit in the approximation of prompt \N decays.
  Also shown are the best upper limits at 95\% \CL from other collider searches in L3~\cite{Achard:2001qv}, DELPHI~\cite{DELPHI},
  ATLAS~\cite{Aad:2015xaa}, and CMS~\cite{Khachatryan:2015gha}.
}
\label{fig:interpr}
\end{figure*}

The interpretation of the results is presented in Fig.~\ref{fig:interpr}.
The \N lifetime is inversely proportional to $\mN^{5} \abs{\VlN}^2$~\cite{Izaguirre:2015pga, Dube:2017jgo}.
At low masses this becomes significant, resulting in displaced decays and lower
efficiency than if the decays were prompt, illustrated by comparison of the black dotted line in Fig.~\ref{fig:interpr}
(prompt assumption) with the final result.
This is accounted for by calculating the efficiency vs. \N lifetime, and propagating this to the limits on mixing parameter vs. mass.

In summary, a search has been performed for a heavy neutral lepton \N of Majorana nature produced in the decays of
a $\PW$ boson, with subsequent prompt decays of \N to $\PW\ell$,
where the vector boson decays to $\ell\nu$.
The event signature consists of three charged leptons in any combination of electrons and muons.
No statistically significant excess of events over the expected standard model background is observed.

Upper limits at 95\% confidence level are set on the mixing parameters $\abs{\VeN}^2$ and $\abs{\VmN}^2$,
ranging between $1.2\times10^{-5}$ and 1.8 for \N masses in the range $1\GeV < \mN < 1.2\TeV$.
These results surpass those obtained in previous searches carried out
by the ATLAS~\cite{Aad:2015xaa} and CMS~\cite{Khachatryan:2015gha,Khachatryan:2016olu} Collaborations, and
are the first direct limits for $\mN> 500\GeV$. This search also provides the first probes for low masses ($\mN< 40\GeV$) at
the LHC, improving on the limits set previously by the L3~\cite{L3} and DELPHI~\cite{DELPHI} Collaborations.
For \N masses below 3\GeV, the most stringent limits to date are obtained from the beam-dump experiments:
CHARM~\cite{Bergsma:1985is,Vilain:1994vg}, BEBC~\cite{CooperSarkar:1985nh},
FMMF~\cite{Gallas:1994xp}, and NuTeV~\cite{Vaitaitis:1999wq}.

\begin{acknowledgments}
\hyphenation{Bundes-ministerium Forschungs-gemeinschaft Forschungs-zentren} We congratulate our colleagues in the CERN accelerator departments for the excellent performance of the LHC and thank the technical and administrative staffs at CERN and at other CMS institutes for their contributions to the success of the CMS effort. In addition, we gratefully acknowledge the computing centers and personnel of the Worldwide LHC Computing Grid for delivering so effectively the computing infrastructure essential to our analyses. Finally, we acknowledge the enduring support for the construction and operation of the LHC and the CMS detector provided by the following funding agencies: BMWFW and FWF (Austria); FNRS and FWO (Belgium); CNPq, CAPES, FAPERJ, and FAPESP (Brazil); MES (Bulgaria); CERN; CAS, MoST, and NSFC (China); COLCIENCIAS (Colombia); MSES and CSF (Croatia); RPF (Cyprus); SENESCYT (Ecuador); MoER, ERC IUT, and ERDF (Estonia); Academy of Finland, MEC, and HIP (Finland); CEA and CNRS/IN2P3 (France); BMBF, DFG, and HGF (Germany); GSRT (Greece); OTKA and NIH (Hungary); DAE and DST (India); IPM (Iran); SFI (Ireland); INFN (Italy); MSIP and NRF (Republic of Korea); LAS (Lithuania); MOE and UM (Malaysia); BUAP, CINVESTAV, CONACYT, LNS, SEP, and UASLP-FAI (Mexico); MBIE (New Zealand); PAEC (Pakistan); MSHE and NSC (Poland); FCT (Portugal); JINR (Dubna); MON, RosAtom, RAS, RFBR and RAEP (Russia); MESTD (Serbia); SEIDI and CPAN (Spain); Swiss Funding Agencies (Switzerland); MST (Taipei); ThEPCenter, IPST, STAR, and NSTDA (Thailand); TUBITAK and TAEK (Turkey); NASU and SFFR (Ukraine); STFC (United Kingdom); DOE and NSF (USA).\par
\end{acknowledgments}

\bibliography{auto_generated}

\ifthenelse{\boolean{cms@external}}{}{
\clearpage
\numberwithin{figure}{section}
\appendix
\section{Expected and observed yields in the search regions\label{app:suppMat}}
\input{supplemental_material}
}
\cleardoublepage \section{The CMS Collaboration \label{app:collab}}\begin{sloppypar}\hyphenpenalty=5000\widowpenalty=500\clubpenalty=5000\input{EXO-17-012-authorlist.tex}\end{sloppypar}
\end{document}

%% file: supplemental_material.tex
\ifthenelse{\boolean{cms@external}}{\providecommand{\figA}{Fig.~1 in the main document}}{\providecommand{\figA}{Fig.~\ref{fig:SRplots}}}

\begin{table*}[tbh]
\centering
\topcaption{Observed (expected) event yields in the low-mass search region.
The uncertainties contain both the statistical and systematic components.}
\label{tab:lowMSRres}
\resizebox{1.0\textwidth}{!}{
\begin{scotch}{cc rr@{}p{3mm}@{}l rr@{}p{3mm}@{}l rr@{}p{3mm}@{}l rr@{}p{3mm}@{}l}
\multirow{2}{*}{Flavor} &  \multirow{2}{*}{$\pt^\text{leading}$ (GeV)} &  \multicolumn{16}{ c } {$\minMOS$  (GeV)} \\[3pt] \cline{3-18}
 &  & \multicolumn{4}{c|}{$ < 10$} & \multicolumn{4}{c|}{10--20} & \multicolumn{4}{c|}{20--30} & \multicolumn{4}{c}{$> 30$} \\
\hline \hline
\multirow{2}{*}{$\Pe^\pm\Pe^\pm\mu^\mp$} & $< 30$ & 1 & (0.61 & $\pm$ & 0.44) & 0 & (0.45 & $\pm$ & 0.48) & 0 & (0.14 & $\pm$ & $^{0.25}_{0.14}$) & 3 & (0.27 & $\pm$ & $^{0.50}_{0.27}$) \\
& 30--55 & 1 & (1.05 & $\pm$ & 0.82) & 1 & (0.28 & $\pm$ & 0.25) & 2 & (0.53 & $\pm$ & 0.47) & 0 & (1.7 & $\pm$ & 1.1) \\
[\cmsTabSkip]
\multirow{2}{*}{$\Pe^\pm\mu^\mp\mu^\mp$} & $< 30$ & 5 & (3.0 & $\pm$ & 1.4) & 3 & (2.6 & $\pm$ & 1.3) & 3 & (1.38 & $\pm$ & 0.77) & 0 & (1.71 & $\pm$ & 0.83) \\
& 30--55 & 3 & (2.5 & $\pm$ & 1.2) & 2 & (2.4 & $\pm$ & 1.2) & 2 & (2.6 & $\pm$ & 1.2) & 3 & (1.0 & $\pm$ & $^{1.9}_{1.0}$) \\
\end{scotch}
}
\end{table*}

\begin{table*}[tbh]
\centering
\topcaption{Observed (expected) event yields in the high-mass search region for events with no OSSF lepton pair.
The uncertainties contain both the statistical and systematic components.}
\label{tab:SRhighmassnoOSSFres}
\resizebox{0.9\textwidth}{!}{
\begin{scotch}{ccc rr@{}p{3mm}@{}l rr@{}p{3mm}@{}l rr@{}p{3mm}@{}l}
\multirow{2}{*}{Flavor} & \multirow{2}{*}{$\Mtril$ (GeV)} & \multirow{2}{*}{$\MT$ (GeV)} & \multicolumn{12}{ c } {$\minMOS$  (GeV)} \\[3pt] \cline{4-15}
& & & \multicolumn{4}{c|}{ $ < 100 $} &  \multicolumn{4}{c|}{100--200} &  \multicolumn{4}{c}{ $ > 200$} \\
\hline \hline
\multirow{6}{*}{$\Pe^\pm\Pe^\pm\mu^\mp$} & \multirow{2}{*}{$ < 100$} & $ < 100$ & \multicolumn{4}{c}{} & 1 & (1.45 & $\pm$ & 0.63) & \multicolumn{4}{c}{} \\
 & & $ > 100$ & \multicolumn{4}{c}{} & 0 & (0.43 & $\pm$ & 0.20) & \multicolumn{4}{c}{} \\ \cline{2-15}
 & \multirow{4}{*}{$ > 100$} & $ < 100 $ & 16 & (12.4 & $\pm$ & 2.7) & 2 & (1.31 & $\pm$ &0.34) & \multirow{4}{*}{1} & \multirow{4}{*}{(0.54} & \multirow{4}{*}{$\pm$} & \multirow{4}{*}{0.16)} \\
 & & 100--150 & 4 & (4.1 & $\pm$ & 10) & \multirow{3}{*}{1} &  \multirow{3}{*}{(1.16} &  \multirow{3}{*}{$\pm$} &  \multirow{3}{*}{0.29)} & & & & \\
 & & 150--250 & 2 & (1.99 & $\pm$ & 0.55) & & & & & & & \\
 & & $ > 250 $ & 1 & (0.70 & $\pm$ & 0.43) & & & & & & & \\
\hline
\multirow{6}{*}{$\Pe^\pm\mu^\mp\mu^\mp$} & \multirow{2}{*}{$ < 100$} & $ < 100$ & \multicolumn{4}{c}{} & 2 & (0.88 & $\pm$ & 0.32) & \multicolumn{4}{c}{} \\
 & & $ > 100$ & \multicolumn{4}{c}{} & 1 & (0.54 & $\pm$ & 0.29) & \multicolumn{4}{c}{} \\ \cline{2-15}
 & \multirow{4}{*}{$ > 100$} & $ < 100 $ & 12 & (10.9 & $\pm$ & 2.3) & 1 & (1.93 & $\pm$ &0.60) & \multirow{4}{*}{0} & \multirow{4}{*}{(0.160} & \multirow{4}{*}{$\pm$} & \multirow{4}{*}{0.093)} \\
 & & 100--150 & 5 & (4.1 & $\pm$ & 1.1) & \multirow{3}{*}{0} &  \multirow{3}{*}{(0.64} &  \multirow{3}{*}{$\pm$} &  \multirow{3}{*}{0.24)} & & & & \\
 & & 150--250 & 2 & (2.72 & $\pm$ & 0.73) & & & & & & & \\
 & & $ > 250 $ & 0 & (0.44 & $\pm$ & 0.19) & & & & & & & \\
\end{scotch}
}
\end{table*}

\begin{table*}[tbh]
\centering
\topcaption{Observed (expected) event yields in the high-mass search region for events with an OSSF lepton pair.
The uncertainties contain both the statistical and systematic components.}
\label{tab:SRhighmassOSSFres}
\resizebox{0.9\textwidth}{!}{
\begin{scotch}{ccc rr@{}p{3mm}@{}l rr@{}p{3mm}@{}l rr@{}p{3mm}@{}l}
\multirow{2}{*}{Flavor} & \multirow{2}{*}{$\Mtril$ (GeV)} & \multirow{2}{*}{$\MT$ (GeV)} & \multicolumn{12}{ c } {$\minMOS$  (GeV)} \\[3pt] \cline{4-15}
& & & \multicolumn{4}{c|}{ $ < 100 $} &  \multicolumn{4}{c|}{100--200} &  \multicolumn{4}{c}{ $ > 200$} \\
\hline \hline
\multirow{8}{*}{$\ge2\Pe$} & \multirow{3}{*}{$ < 100$} & $ < 100$ & \multicolumn{4}{c}{} & 10 & (12.3 & $\pm$ & 1.7) & \multicolumn{4}{c}{} \\
 & & 100--200 & \multicolumn{4}{c}{} & 3 & (1.67 & $\pm$ & 0.35) & \multicolumn{4}{c}{} \\
 & & $ > 200$ & \multicolumn{4}{c}{} & 0 & (0.226 & $\pm$ & 0.064) & \multicolumn{4}{c}{} \\ \cline{2-15}
 & \multirow{5}{*}{$ > 100$} & $ < 100 $ & 127 & (131 & $\pm$ & 14) & 31 & (38.2 & $\pm$ &4.3) & 6 & (4.56 & $\pm$ & 0.94) \\
 & & 100--200 & 34 & (40.9 & $\pm$ & 4.9) & 8 & (12.7 & $\pm$ &1.8) & 0 & (2.37 & $\pm$ & 0.50) \\
 & & 200--300 & 3 & (5.28 & $\pm$ & 0.78) & 1 & (1.81 & $\pm$ &0.29) & 0 & (0.57 & $\pm$ & 0.13) \\
 & & 300--400 & 1 & (1.20 & $\pm$ & 0.24) & \multirow{2}{*}{0} & \multirow{2}{*}{(0.86} & \multirow{2}{*}{$\pm$} & \multirow{2}{*}{0.17)} & \multirow{2}{*}{0} & \multirow{2}{*}{(0.61} & \multirow{2}{*}{$\pm$} & \multirow{2}{*}{0.14)} \\
 & & $ > 400 $ & 1 & (0.87 & $\pm$ & 0.24) & & & & & & & & \\
\hline
\multirow{8}{*}{$\ge2\mu$} & \multirow{3}{*}{$ < 100$} & $ < 100$ & \multicolumn{4}{c}{} & 30 & (24.4 & $\pm$ & 2.9) & \multicolumn{4}{c}{} \\
 & & 100--200 & \multicolumn{4}{c}{} & 3 & (3.64 & $\pm$ & 0.56) & \multicolumn{4}{c}{} \\
 & & $ > 200$ & \multicolumn{4}{c}{} & 1 & (0.63 & $\pm$ & 0.22) & \multicolumn{4}{c}{} \\ \cline{2-15}
 & \multirow{5}{*}{$ > 100$} & $ < 100 $ & 220 & (217 & $\pm$ & 22) & 63 & (65.6 & $\pm$ &6.6) & 8 & (6.6 & $\pm$ & 1.4) \\
 & & 100--200 & 77 & (61.3 & $\pm$ & 6.9) & 23 & (17.6 & $\pm$ &2.3) & 7 & (3.50 & $\pm$ & 0.63) \\
 & & 200--300 & 8 & (9.1 & $\pm$ & 1.2) & 1 & (3.08 & $\pm$ &0.54) & 0 & (1.64 & $\pm$ & 0.73) \\
 & & 300--400 & 1 & (2.10 & $\pm$ & 0.44) & \multirow{2}{*}{1} & \multirow{2}{*}{(1.34} & \multirow{2}{*}{$\pm$} & \multirow{2}{*}{0.24)} & \multirow{2}{*}{0} & \multirow{2}{*}{(0.62} & \multirow{2}{*}{$\pm$} & \multirow{2}{*}{0.59)} \\
 & & $ > 400 $ & 2 & (1.54 & $\pm$ & 0.30) & & & & & & & & \\
\end{scotch}
}
\end{table*}

\begin{figure*}[htb]
\centering
  \includegraphics[width=0.49\textwidth]{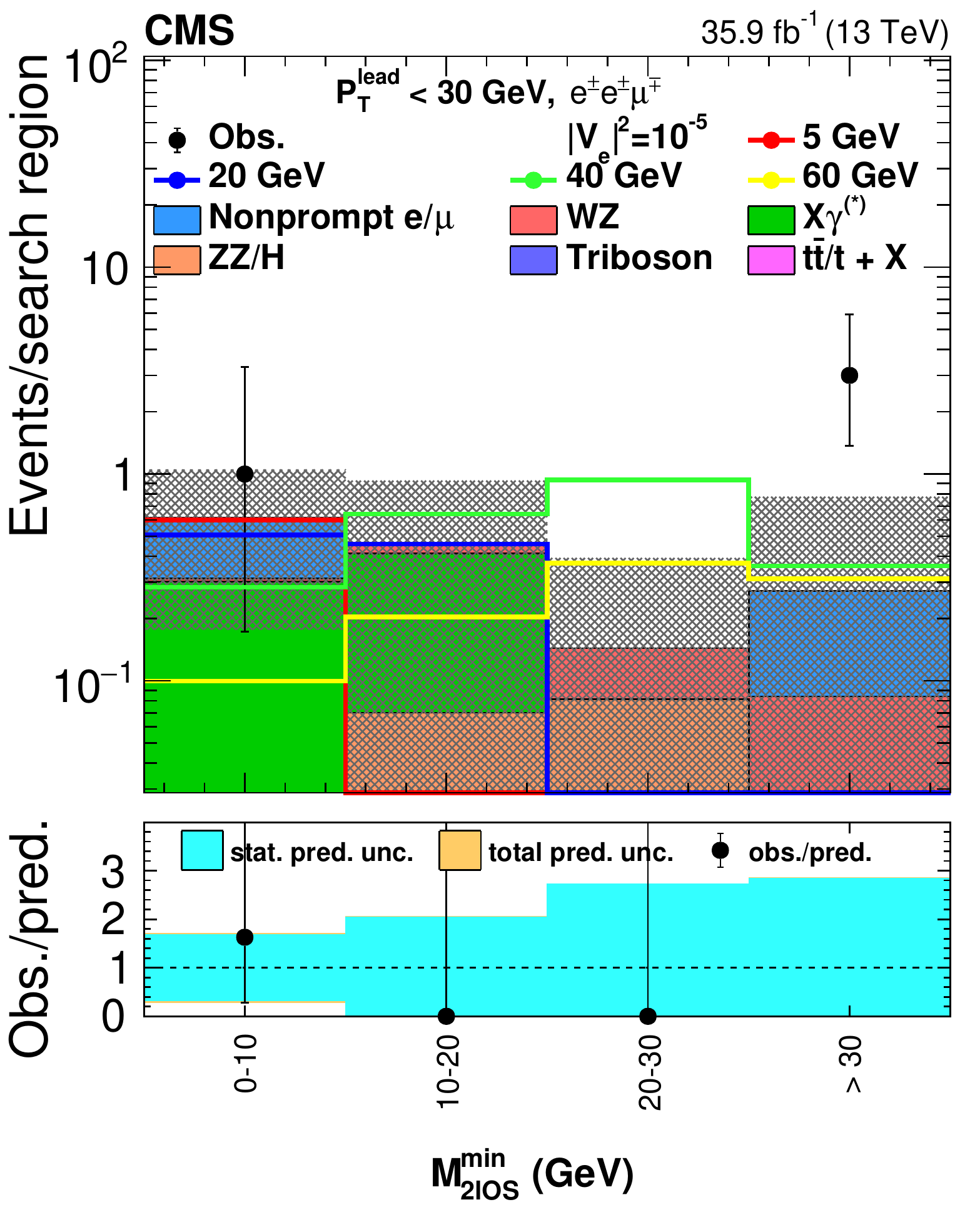}
  \includegraphics[width=0.49\textwidth]{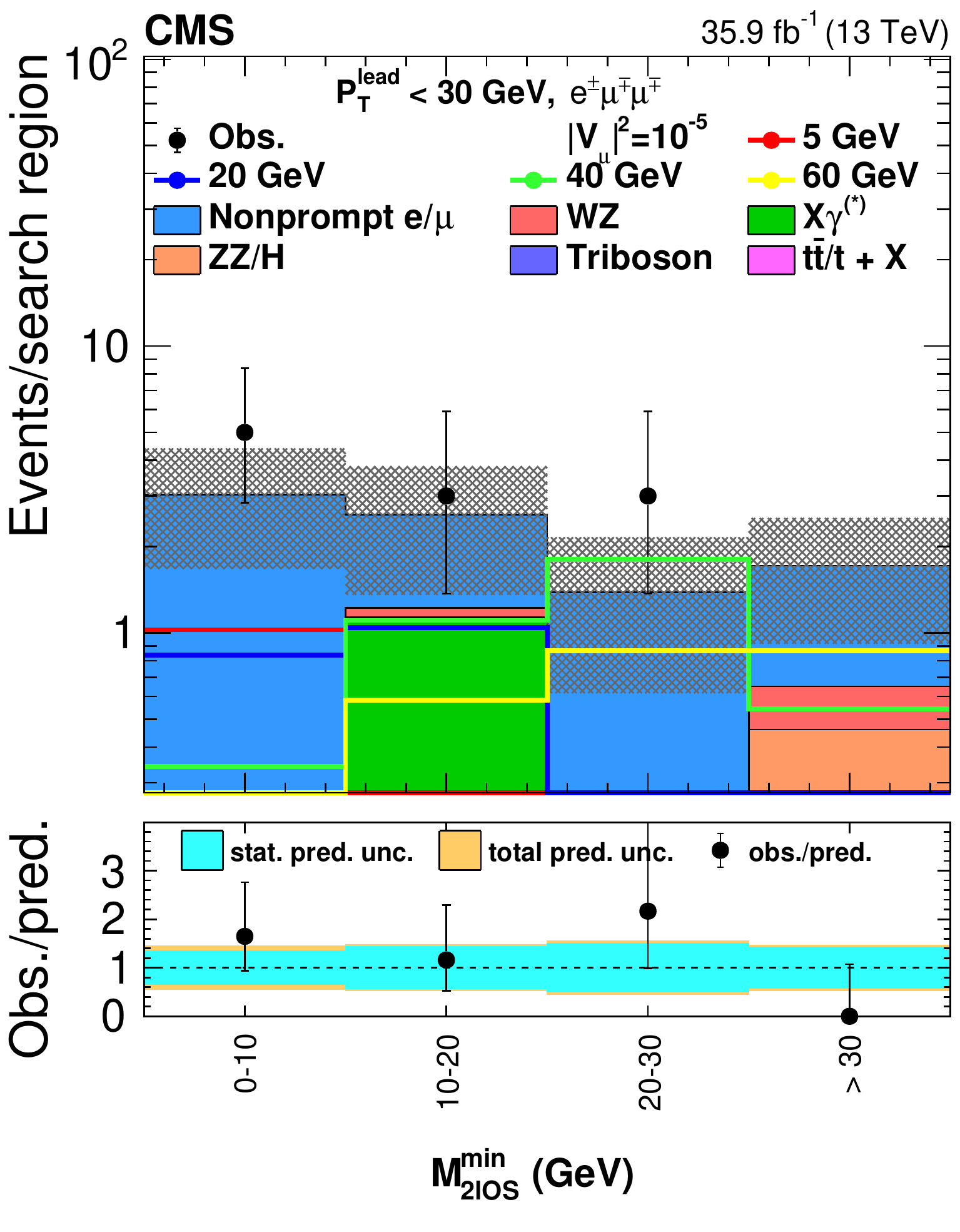}
\caption{Expanded version of \figA, 
showing the observed and expected event yields in the low-mass region with $\pt^\text{leading} < 30\GeV$, for events with at least 2 electrons (left) and 2 muons (right).  Contributions from various possible signals are shown for comparison.}
\label{fig:lowM_3lnoOSSF_lowPt}
\end{figure*}

\begin{figure*}[htb]
\centering
  \includegraphics[width=0.49\textwidth]{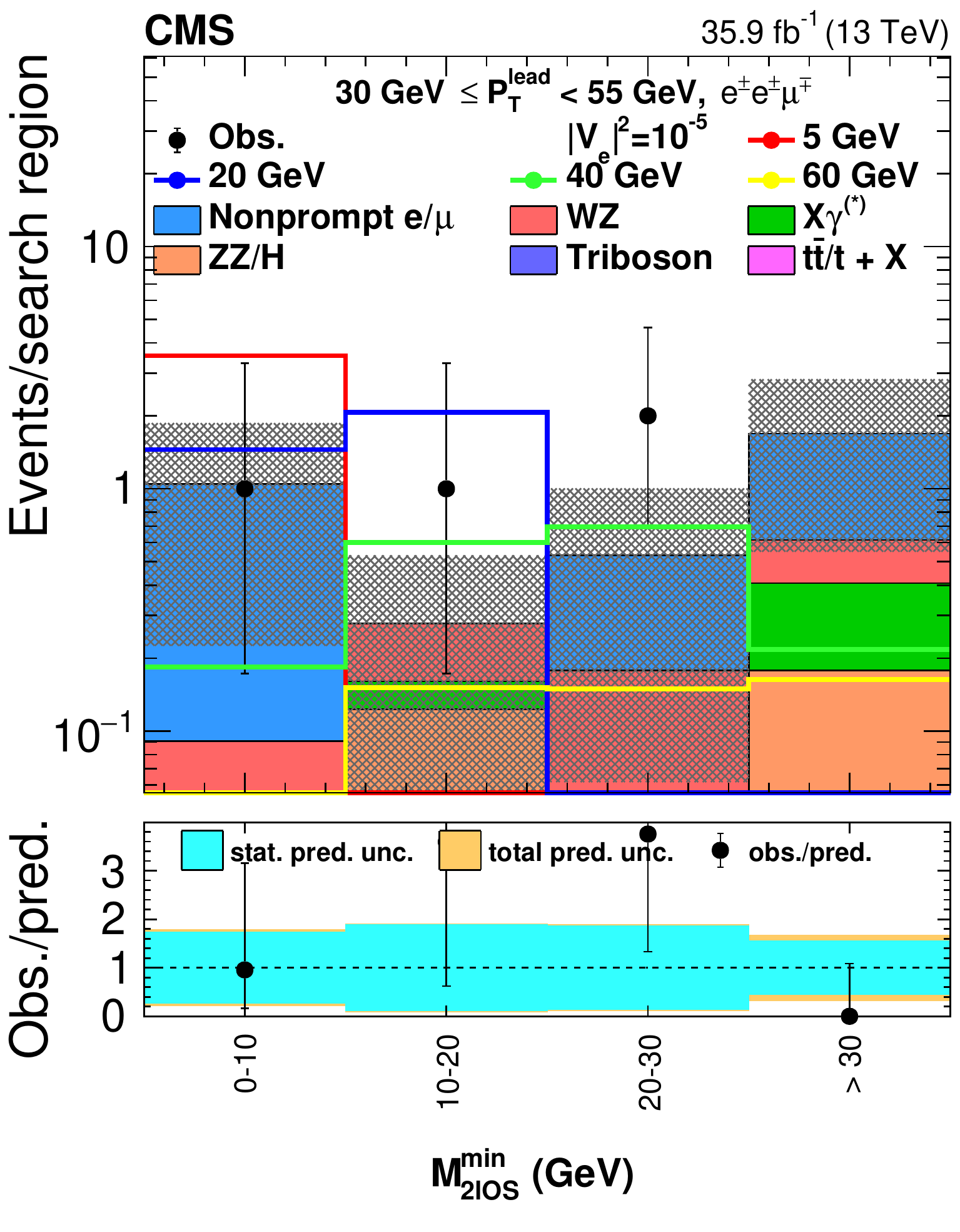}
  \includegraphics[width=0.49\textwidth]{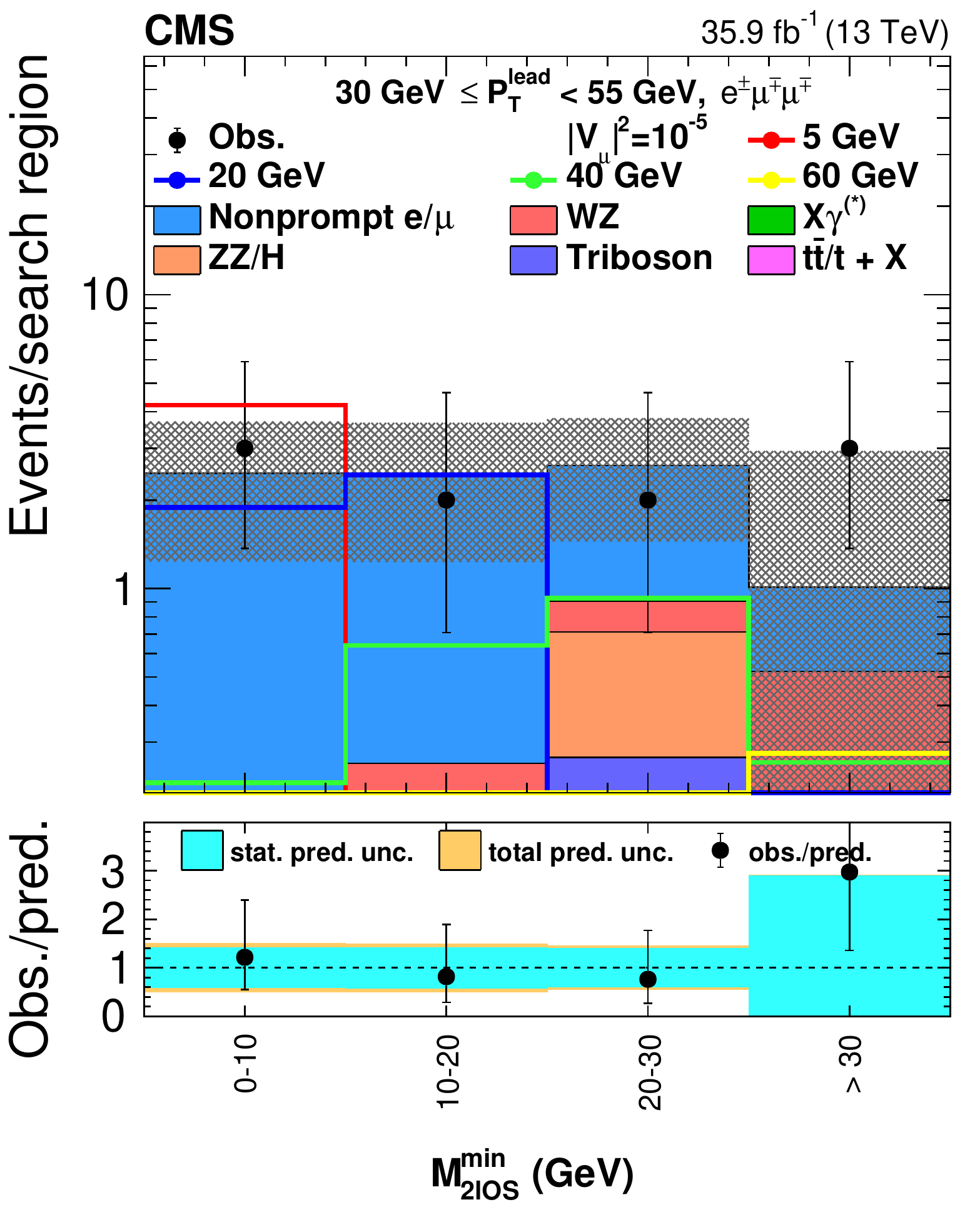}
\caption{Expanded version of \figA, 
showing the observed and expected event yields in the low-mass region with $30 \GeV \leq \pt^\text{leading} < 55\GeV$, for events with at least 2 electrons (left) and 2 muons (right). Contributions from various possible signals are shown for comparison.}
\label{fig:lowM_3lnoOSSF_highPt}
\end{figure*}

\begin{figure*}[htb]
\centering
  \includegraphics[width=0.49\textwidth]{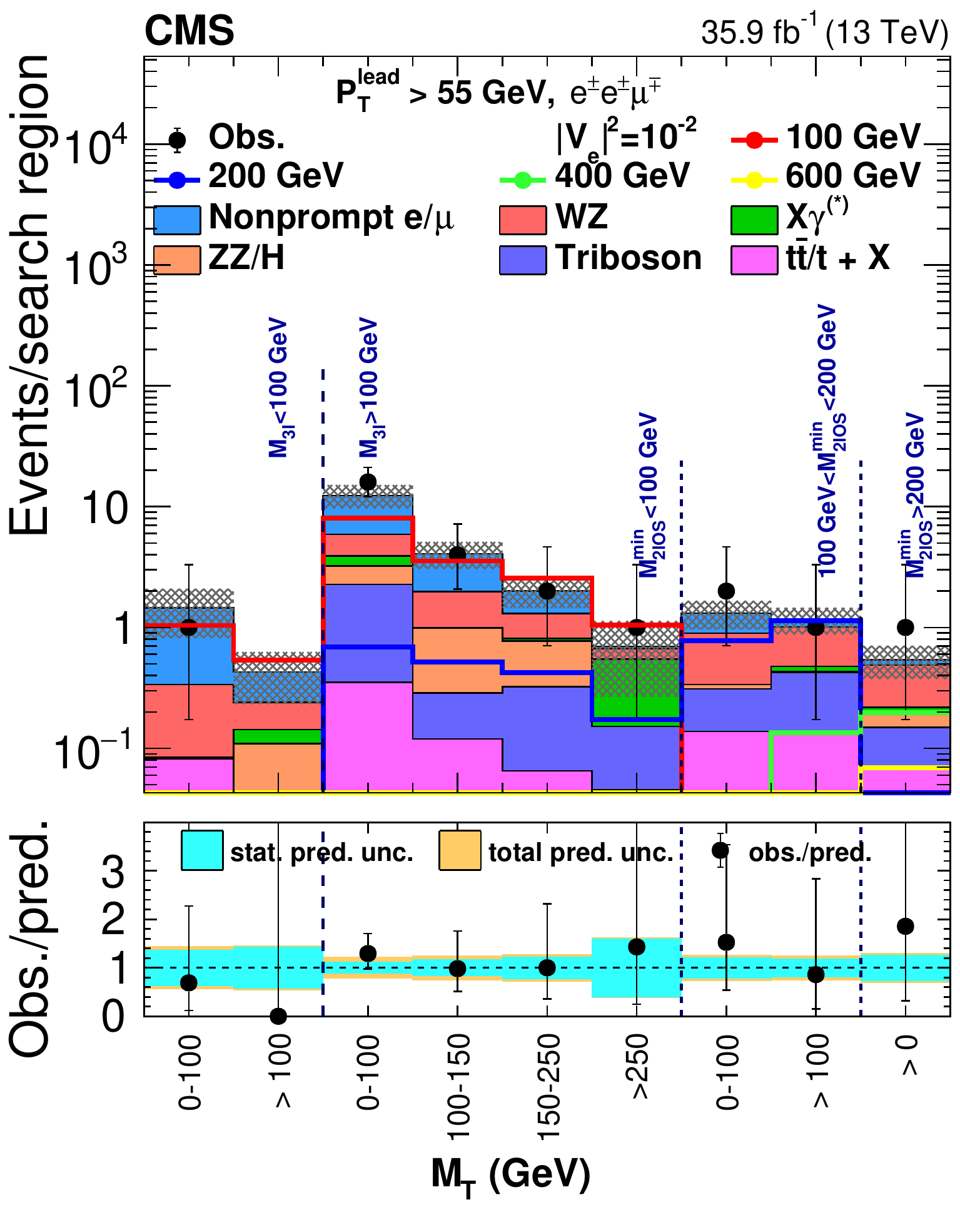}
  \includegraphics[width=0.49\textwidth]{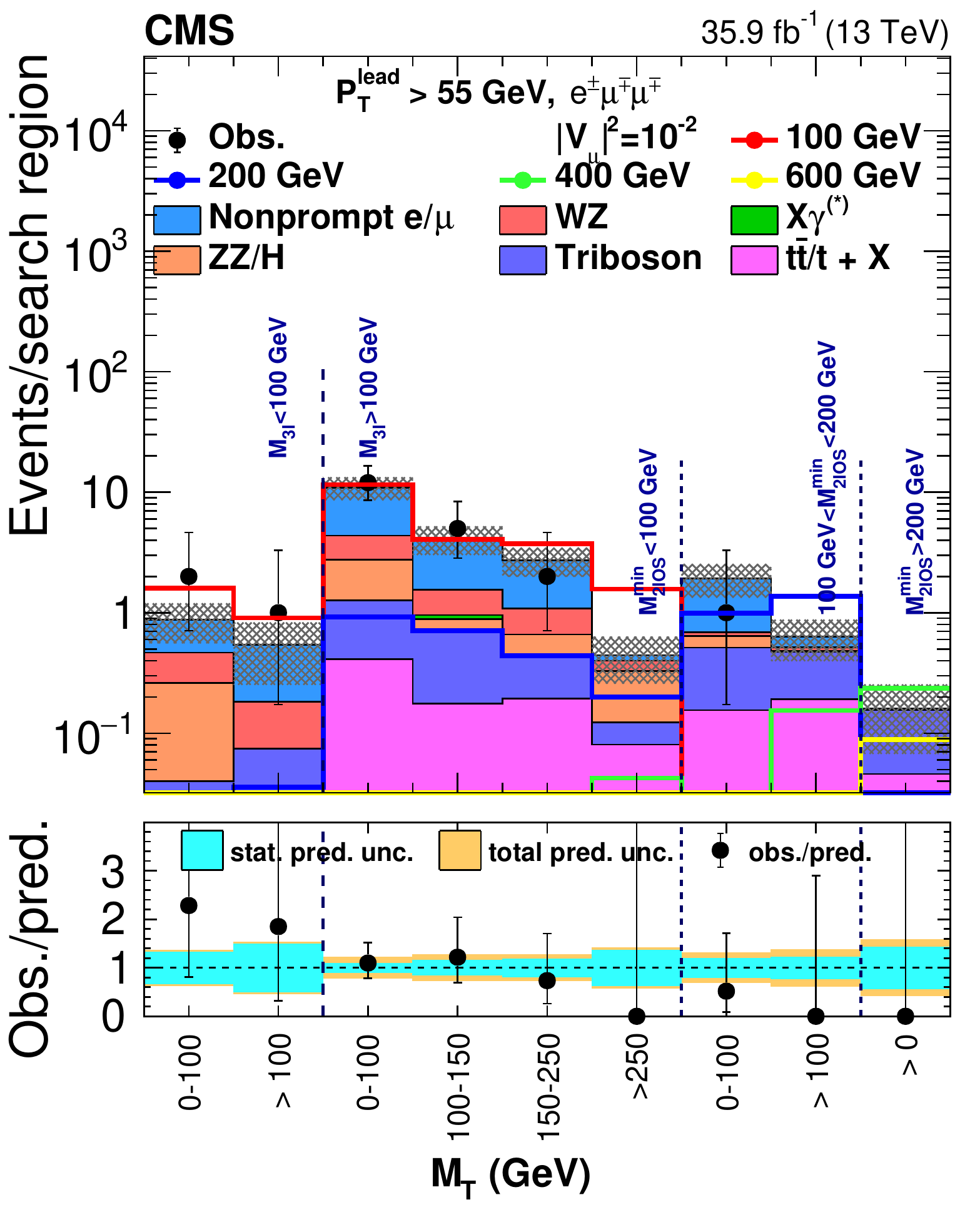}
\caption{Expanded version of \figA, 
showing the observed and expected event yields in the high-mass region without an OSSF pair, for events with at least 2 electrons (left) and 2 muons (right). Contributions from various possible signals are shown for comparison.}
\label{fig:highM_3lnoOSSF}
\end{figure*}

\begin{figure*}[htb]
\centering
  \includegraphics[width=0.49\textwidth]{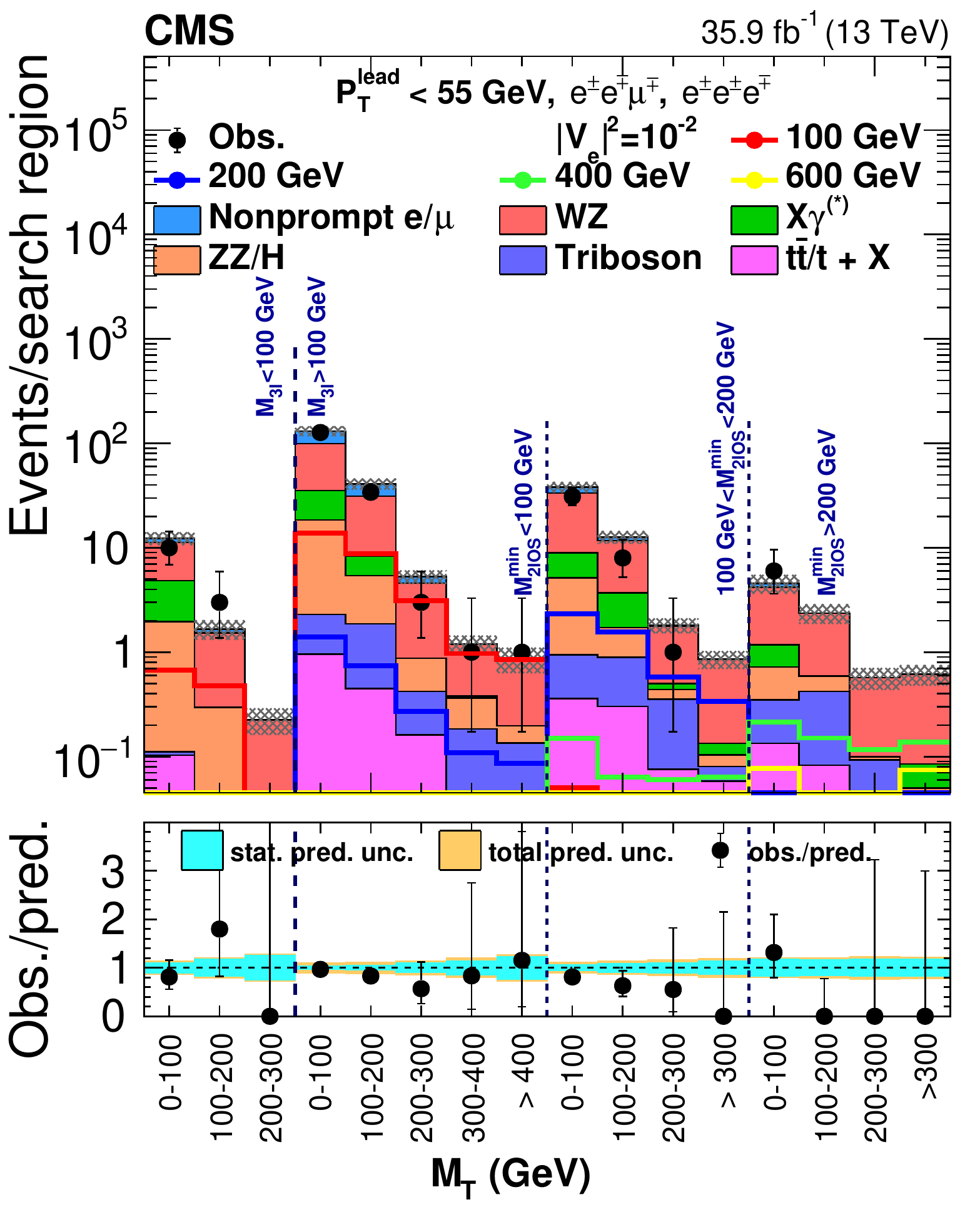}
  \includegraphics[width=0.49\textwidth]{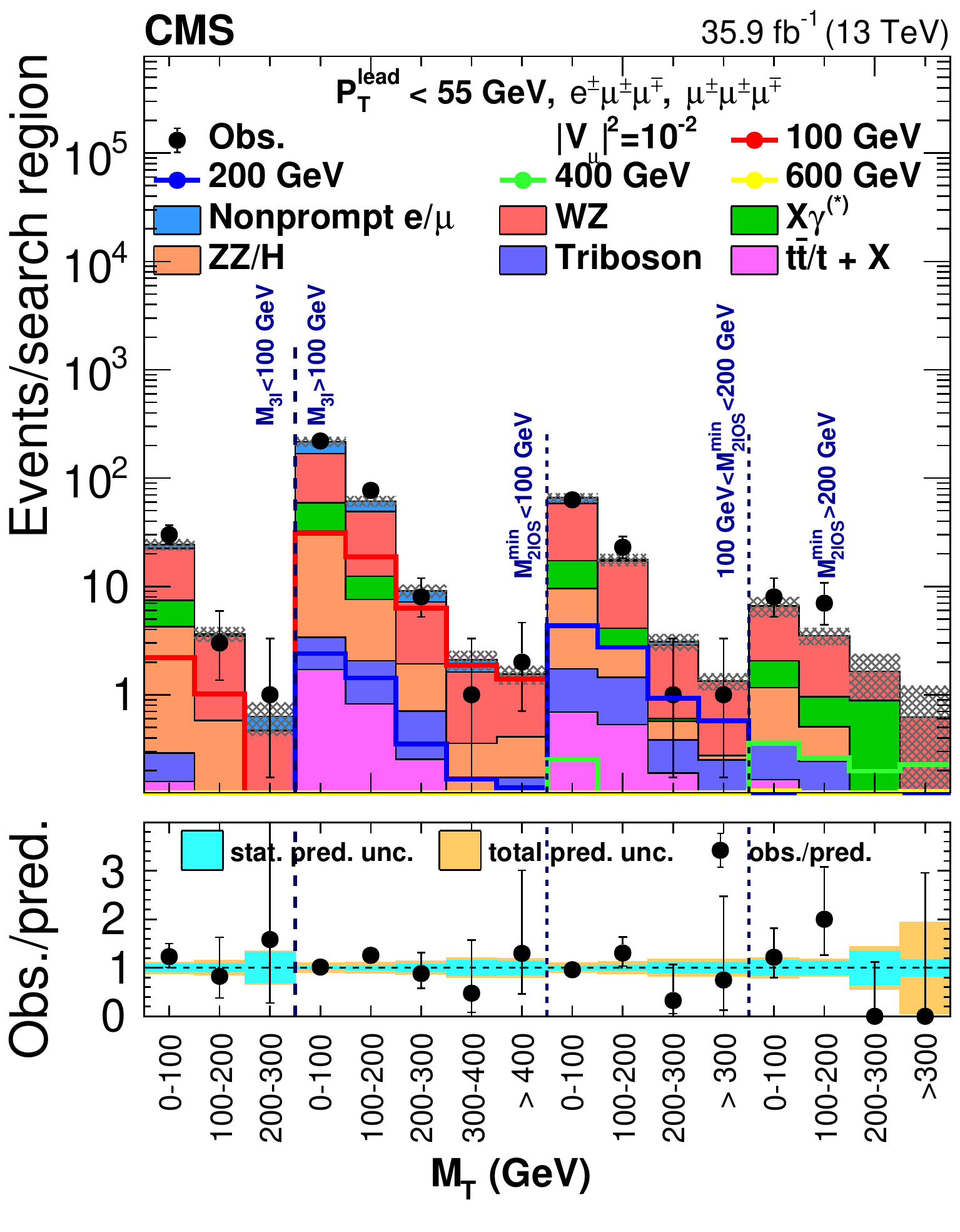}
\caption{Expanded version of \figA, 
showing the observed and expected event yields in the high-mass region with an OSSF pair, for events with at least 2 electrons (left) and 2 muons (right). Contributions from various possible signals are shown for comparison.}
\label{fig:highM_3lOSSF}
\end{figure*}

%% file: EXO-17-012-authorlist.tex
\textbf{Yerevan Physics Institute,  Yerevan,  Armenia}\\*[0pt]
A.M.~Sirunyan, A.~Tumasyan
\vskip\cmsinstskip
\textbf{Institut f\"{u}r Hochenergiephysik,  Wien,  Austria}\\*[0pt]
W.~Adam, F.~Ambrogi, E.~Asilar, T.~Bergauer, J.~Brandstetter, E.~Brondolin, M.~Dragicevic, J.~Er\"{o}, A.~Escalante Del Valle, M.~Flechl, M.~Friedl, R.~Fr\"{u}hwirth\cmsAuthorMark{1}, V.M.~Ghete, J.~Grossmann, J.~Hrubec, M.~Jeitler\cmsAuthorMark{1}, A.~K\"{o}nig, N.~Krammer, I.~Kr\"{a}tschmer, D.~Liko, T.~Madlener, I.~Mikulec, E.~Pree, N.~Rad, H.~Rohringer, J.~Schieck\cmsAuthorMark{1}, R.~Sch\"{o}fbeck, M.~Spanring, D.~Spitzbart, A.~Taurok, W.~Waltenberger, J.~Wittmann, C.-E.~Wulz\cmsAuthorMark{1}, M.~Zarucki
% ...
\vskip\cmsinstskip
% ...
1:~~Also at Vienna University of Technology, Vienna, Austria\\
% ...


%% file: EXO-17-012_temp.bbl
\providecommand{\href}[2]{#2}\begingroup\raggedright\begin{thebibliography}{10}%
\makeatletter
\providecommand{\hrefCMSnoop }[0]{\@secondoftwo}%
\makeatother
\providecommand{\doi}{\texttt{doi:}\begingroup \urlstyle{tt}\Url}

\bibitem{seesawI_1}
\hrefCMSnoop {}{P.~Minkowski, ``{$\mu \to \mathrm{e} \gamma$ at a rate of one
  out of 1-billion muon decays?}'',} \textit{ Phys. Lett. B} \textbf{ 67}
  (1977) 421,
\href{http://dx.doi.org/10.1016/0370-2693(77)90435-X}{\doi{10.1016/0370-2693(77)90435-X}}.

\bibitem{seesawI_2}
\hrefCMSnoop {}{M.~Gell-Mann, P.~Ramond, and R.~Slansky} in \textit{
  {Supergravity: proceedings of the Supergravity Workshop at Stony Brook}},
  P.~V. Nieuwenhuizen and D.~Z. Freedman, eds.
\newblock North-holland, Amsterdam, Netherlands, 1979.

\bibitem{seesawI_3}
\hrefCMSnoop {}{T.~Yanagida} in \textit{ {Proceedings: Workshop on the Unified
  Theories and the Baryon Number in the Universe}}, O.~Sawada and A.~Sugamoto,
  eds.
\newblock Natl. Lab. High Energy Phys., Tsukuba, Japan,
1979.
\newblock

\bibitem{seesawI_4}
\hrefCMSnoop {}{R.~N. Mohapatra and G.~Senjanovi{\'c}, ``{Neutrino mass and
  spontaneous parity violation}'',} \textit{ Phys. Rev. Lett.} \textbf{ 44}
  (1980) 912,
\href{http://dx.doi.org/10.1103/PhysRevLett.44.912}{\doi{10.1103/PhysRevLett.44.912}}.

\bibitem{seesawII_1}
\hrefCMSnoop {}{M.~Magg and C.~Wetterich, ``{Neutrino mass problem and gauge
  hierarchy}'',} \textit{ Phys. Lett. B} \textbf{ 94} (1980) 61,
  \href{http://dx.doi.org/10.1016/0370-2693(80)90825-4}{\doi{10.1016/0370-2693(80)90825-4}}.

\bibitem{seesawII_2}
\hrefCMSnoop {}{J.~Schechter and J.~W.~F. Valle, ``{Neutrino masses in
  $\mathrm{ SU(2) {\otimes} U(1) }$ theories}'',} \textit{ Phys. Rev. D}
  \textbf{ 22} (1980) 2227,
  \href{http://dx.doi.org/10.1103/PhysRevD.22.2227}{\doi{10.1103/PhysRevD.22.2227}}.

\bibitem{seesawII_3}
\hrefCMSnoop {}{T.~P. Cheng and L.-F. Li, ``{Neutrino masses, mixings and
  oscillations in $\mathrm{ SU(2) {\otimes} U(1) }$ models of electroweak
  interactions}'',} \textit{ Phys. Rev. D} \textbf{ 22} (1980) 2860,
\href{http://dx.doi.org/10.1103/PhysRevD.22.2860}{\doi{10.1103/PhysRevD.22.2860}}.

\bibitem{seesawII_4}
\hrefCMSnoop {}{G.~Lazarides, Q.~Shafi, and C.~Wetterich, ``{Proton lifetime
  and fermion masses in an SO(10) model}'',} \textit{ Nucl. Phys. B} \textbf{
  181} (1981) 287,
\href{http://dx.doi.org/10.1016/0550-3213(81)90354-0}{\doi{10.1016/0550-3213(81)90354-0}}.

\bibitem{seesawII_5}
\hrefCMSnoop {}{R.~N. Mohapatra and G.~Senjanovi{\'c}, ``Neutrino masses and
  mixings in gauge models with spontaneous parity violation'',} \textit{ Phys.
  Rev. D} \textbf{ 23} (1981) 165,
  \href{http://dx.doi.org/10.1103/PhysRevD.23.165}{\doi{10.1103/PhysRevD.23.165}}.

\bibitem{seesawII_6}
\hrefCMSnoop {}{J.~Schechter and J.~W.~F. Valle, ``Neutrino decay and
  spontaneous violation of lepton number'',} \textit{ Phys. Rev. D} \textbf{
  25} (1982) 774,
  \href{http://dx.doi.org/10.1103/PhysRevD.25.774}{\doi{10.1103/PhysRevD.25.774}}.

\bibitem{seesawIII}
\hrefCMSnoop {}{R.~Foot, H.~Lew, X.-G. He, and G.~C. Joshi, ``See-saw neutrino
  masses induced by a triplet of leptons'',} \textit{ Z. Phys. C} \textbf{ 44}
  (1989) 441,
  \href{http://dx.doi.org/10.1007/BF01415558}{\doi{10.1007/BF01415558}}.

\bibitem{Appelquist:2002me}
\hrefCMSnoop {}{T.~Appelquist and R.~Shrock, ``{Neutrino masses in theories
  with dynamical electroweak symmetry breaking}'',} \textit{ Phys. Lett. B}
  \textbf{ 548} (2002) 204,
  \href{http://dx.doi.org/10.1016/S0370-2693(02)02854-X}{\doi{10.1016/S0370-2693(02)02854-X}},
\href{http://www.arXiv.org/abs/hep-ph/0204141}{\texttt{arXiv:hep-ph/0204141}}.

\bibitem{Appelquist:2003uu}
\hrefCMSnoop {}{T.~Appelquist and R.~Shrock, ``{Dynamical symmetry breaking of
  extended gauge symmetries}'',} \textit{ Phys. Rev. Lett.} \textbf{ 90} (2003)
  201801,
  \href{http://dx.doi.org/10.1103/PhysRevLett.90.201801}{\doi{10.1103/PhysRevLett.90.201801}},
\href{http://www.arXiv.org/abs/hep-ph/0301108}{\texttt{arXiv:hep-ph/0301108}}.

\bibitem{Asaka:2005an}
\hrefCMSnoop {}{T.~Asaka, S.~Blanchet, and M.~Shaposhnikov, ``{The $\nu$MSM,
  dark matter and neutrino masses}'',} \textit{ Phys. Lett. B} \textbf{ 631}
  (2005) 151,
  \href{http://dx.doi.org/10.1016/j.physletb.2005.09.070}{\doi{10.1016/j.physletb.2005.09.070}},
\href{http://www.arXiv.org/abs/hep-ph/0503065}{\texttt{arXiv:hep-ph/0503065}}.

\bibitem{Asaka:2005pn}
\hrefCMSnoop {}{T.~Asaka and M.~Shaposhnikov, ``{The $\nu$MSM, dark matter and
  baryon asymmetry of the universe}'',} \textit{ Phys. Lett. B} \textbf{ 620}
  (2005) 17,
  \href{http://dx.doi.org/10.1016/j.physletb.2005.06.020}{\doi{10.1016/j.physletb.2005.06.020}},
\href{http://www.arXiv.org/abs/hep-ph/0505013}{\texttt{arXiv:hep-ph/0505013}}.

\bibitem{vMSM_lg}
\hrefCMSnoop {}{M.~Fukugita and T.~Yanagida, ``{Baryogenesis without grand
  unification}'',} \textit{ Phys. Lett. B} \textbf{ 174} (1986) 45,
\href{http://dx.doi.org/10.1016/0370-2693(86)91126-3}{\doi{10.1016/0370-2693(86)91126-3}}.

\bibitem{vMSM_no}
\hrefCMSnoop {}{E.~K. Akhmedov, V.~A. Rubakov, and A.~{\relax Yu}. Smirnov,
  ``{Baryogenesis via Neutrino Oscillations}'',} \textit{ Phys. Rev. Lett.}
  \textbf{ 81} (1998) 1359,
  \href{http://dx.doi.org/10.1103/PhysRevLett.81.1359}{\doi{10.1103/PhysRevLett.81.1359}},
\href{http://www.arXiv.org/abs/hep-ph/9803255}{\texttt{arXiv:hep-ph/9803255}}.

\bibitem{Canetti:2012kh}
\hrefCMSnoop {}{L.~Canetti, M.~Drewes, T.~Frossard, and M.~Shaposhnikov,
  ``{Dark matter, baryogenesis and neutrino oscillations from right-handed
  neutrinos}'',} \textit{ Phys. Rev. D} \textbf{ 87} (2013) 093006,
  \href{http://dx.doi.org/10.1103/PhysRevD.87.093006}{\doi{10.1103/PhysRevD.87.093006}},
\href{http://www.arXiv.org/abs/1208.4607}{\texttt{arXiv:1208.4607}}.

\bibitem{Canetti:2014dka}
\hrefCMSnoop {}{L.~Canetti, M.~Drewes, and B.~Garbrecht, ``{Probing
  leptogenesis with GeV-scale sterile neutrinos at LHCb and Belle II}'',}
  \textit{ Phys. Rev. D} \textbf{ 90} (2014) 125005,
  \href{http://dx.doi.org/10.1103/PhysRevD.90.125005}{\doi{10.1103/PhysRevD.90.125005}},
\href{http://www.arXiv.org/abs/1404.7114}{\texttt{arXiv:1404.7114}}.

\bibitem{Antusch:2017pkq}
S.~Antusch\hrefCMSnoop {}{ {et~al.}, ``{Probing Leptogenesis at Future
  Colliders}'',} (2017).
\href{http://www.arXiv.org/abs/1710.03744}{\texttt{arXiv:1710.03744}}.

\bibitem{Evans:2008zzb}
\hrefCMSnoop {}{L.~Evans and P.~Bryant, ``{LHC Machine}'',} \textit{ JINST}
  \textbf{ 3} (2008) S08001,
\href{http://dx.doi.org/10.1088/1748-0221/3/08/S08001}{\doi{10.1088/1748-0221/3/08/S08001}}.

\bibitem{Aad:2011vj}
\hrefCMSnoop {}{{ATLAS Collaboration}, ``{Inclusive search for same-sign
  dilepton signatures in pp collisions at $\sqrt{s}=7$ TeV with the ATLAS
  detector}'',} \textit{ JHEP} \textbf{ 10} (2011) 107,
  \href{http://dx.doi.org/10.1007/JHEP10(2011)107}{\doi{10.1007/JHEP10(2011)107}},
\href{http://www.arXiv.org/abs/1108.0366}{\texttt{arXiv:1108.0366}}.

\bibitem{ATLAS:2012ak}
\hrefCMSnoop {}{{ATLAS Collaboration}, ``{Search for heavy neutrinos and
  right-handed W bosons in events with two leptons and jets in pp collisions at
  $\sqrt{s}=7$ TeV with the ATLAS detector}'',} \textit{ Eur. Phys. J. C}
  \textbf{ 72} (2012) 2056,
  \href{http://dx.doi.org/10.1140/epjc/s10052-012-2056-4}{\doi{10.1140/epjc/s10052-012-2056-4}},
\href{http://www.arXiv.org/abs/1203.5420}{\texttt{arXiv:1203.5420}}.

\bibitem{Chatrchyan:2012fla}
\hrefCMSnoop {}{{CMS Collaboration}, ``{Search for heavy Majorana neutrinos in
  $\mu^{\pm}\mu^{\pm}$ + jets and $\Pe^\pm\Pe^\pm$ + jets events in pp
  collisions at $\sqrt{s} = 7\TeV$}'',} \textit{ Phys. Lett. B} \textbf{ 717}
  (2012) 109,
  \href{http://dx.doi.org/10.1016/j.physletb.2012.09.012}{\doi{10.1016/j.physletb.2012.09.012}},
\href{http://www.arXiv.org/abs/1207.6079}{\texttt{arXiv:1207.6079}}.

\bibitem{CMS:2012zv}
\hrefCMSnoop {}{{CMS Collaboration}, ``{Search for heavy neutrinos and
  W$_\text{R}$ bosons with right-handed couplings in a left-right symmetric
  model in pp collisions at sqrt(s) = 7 TeV}'',} \textit{ Phys. Rev. Lett.}
  \textbf{ 109} (2012) 261802,
  \href{http://dx.doi.org/10.1103/PhysRevLett.109.261802}{\doi{10.1103/PhysRevLett.109.261802}},
\href{http://www.arXiv.org/abs/1210.2402}{\texttt{arXiv:1210.2402}}.

\bibitem{Khachatryan:2014dka}
\hrefCMSnoop {}{{CMS Collaboration}, ``{Search for heavy neutrinos and W bosons
  with right-handed couplings in proton-proton collisions at $\sqrt{s} =
  8\,\text {TeV} $}'',} \textit{ Eur. Phys. J. C} \textbf{ 74} (2014) 3149,
  \href{http://dx.doi.org/10.1140/epjc/s10052-014-3149-z}{\doi{10.1140/epjc/s10052-014-3149-z}},
\href{http://www.arXiv.org/abs/1407.3683}{\texttt{arXiv:1407.3683}}.

\bibitem{Khachatryan:2015gha}
\hrefCMSnoop {}{{CMS Collaboration}, ``{Search for heavy Majorana neutrinos in
  $\mu^\pm \mu^\pm$ + jets events in proton-proton collisions at $\sqrt{s} =
  8\TeV$}'',} \textit{ Phys. Lett. B} \textbf{ 748} (2015) 144,
  \href{http://dx.doi.org/10.1016/j.physletb.2015.06.070}{\doi{10.1016/j.physletb.2015.06.070}},
\href{http://www.arXiv.org/abs/1501.05566}{\texttt{arXiv:1501.05566}}.

\bibitem{Aad:2015xaa}
\hrefCMSnoop {}{{ATLAS Collaboration}, ``{Search for heavy Majorana neutrinos
  with the ATLAS detector in pp collisions at $\sqrt{s}=8\TeV$ }'',} \textit{
  JHEP} \textbf{ 07} (2015) 162,
  \href{http://dx.doi.org/10.1007/JHEP07(2015)162}{\doi{10.1007/JHEP07(2015)162}},
\href{http://www.arXiv.org/abs/1506.06020}{\texttt{arXiv:1506.06020}}.

\bibitem{Khachatryan:2016olu}
\hrefCMSnoop {}{{CMS Collaboration}, ``{Search for heavy Majorana neutrinos in
  $\Pe^\pm\Pe^\pm$ + jets and $\Pe^\pm\mu^\pm$ + jets events in proton-proton
  collisions at $\sqrt{s}=8\TeV$ }'',} \textit{ JHEP} \textbf{ 04} (2016) 169,
  \href{http://dx.doi.org/10.1007/JHEP04(2016)169}{\doi{10.1007/JHEP04(2016)169}},
\href{http://www.arXiv.org/abs/1603.02248}{\texttt{arXiv:1603.02248}}.

\bibitem{CooperSarkar:1985nh}
\hrefCMSnoop {}{{WA66} Collaboration, ``{Search for heavy neutrino decays in
  the BEBC beam dump experiment}'',} \textit{ Phys. Lett. B} \textbf{ 160}
  (1985) 207,
\href{http://dx.doi.org/10.1016/0370-2693(85)91493-5}{\doi{10.1016/0370-2693(85)91493-5}}.

\bibitem{Bergsma:1985is}
\hrefCMSnoop {}{{CHARM} Collaboration, ``{A search for decays of heavy
  neutrinos in the mass range 0.5 -- 2.8 GeV}'',} \textit{ Phys. Lett. B}
  \textbf{ 166} (1986) 473,
\href{http://dx.doi.org/10.1016/0370-2693(86)91601-1}{\doi{10.1016/0370-2693(86)91601-1}}.

\bibitem{Badier:1985wg}
\hrefCMSnoop {}{{NA3} Collaboration, ``{Direct photon production from pions and
  protons at 200 GeV/c}'',} \textit{ Z. Phys. C} \textbf{ 31} (1986) 341,
\href{http://dx.doi.org/10.1007/BF01588030}{\doi{10.1007/BF01588030}}.

\bibitem{Bernardi:1987ek}
G.~Bernardi\hrefCMSnoop {}{ {et~al.}, ``{Further limits on heavy neutrino
  couplings}'',} \textit{ Phys. Lett. B} \textbf{ 203} (1988) 332,
\href{http://dx.doi.org/10.1016/0370-2693(88)90563-1}{\doi{10.1016/0370-2693(88)90563-1}}.

\bibitem{L3}
\hrefCMSnoop {}{{L3} Collaboration, ``{Search for isosinglet neutral heavy
  leptons in Z$^0$ decays}'',} \textit{ Phys. Lett. B} \textbf{ 295} (1992)
  371,
\href{http://dx.doi.org/10.1016/0370-2693(92)91579-X}{\doi{10.1016/0370-2693(92)91579-X}}.

\bibitem{Baranov:1992vq}
\hrefCMSnoop {}{S.~A. Baranov {et~al.}, ``{Search for heavy neutrinos at the
  IHEP-JINR Neutrino Detector}'',} \textit{ Phys. Lett. B} \textbf{ 302} (1993)
  336,
\href{http://dx.doi.org/10.1016/0370-2693(93)90405-7}{\doi{10.1016/0370-2693(93)90405-7}}.

\bibitem{Vilain:1994vg}
\hrefCMSnoop {}{{CHARM II} Collaboration, ``{Search for heavy isosinglet
  neutrinos}'',} \textit{ Phys. Lett. B} \textbf{ 343} (1995) 453,
  \href{http://dx.doi.org/10.1016/0370-2693(94)01422-9}{\doi{10.1016/0370-2693(94)01422-9}}.
[Erratum: {\it Phys. Lett. B} {\bf 351} (1995) 387,
  \DOI{10.1016/0370-2693(94)00440-I}].

\bibitem{Gallas:1994xp}
\hrefCMSnoop {}{{FMMF} Collaboration, ``{Search for neutral weakly interacting
  massive particles in the Fermilab Tevatron wideband neutrino beam}'',}
  \textit{ Phys. Rev. D} \textbf{ 52} (1995) 6,
\href{http://dx.doi.org/10.1103/PhysRevD.52.6}{\doi{10.1103/PhysRevD.52.6}}.

\bibitem{DELPHI}
\hrefCMSnoop {}{{DELPHI} Collaboration, ``{Search for neutral heavy leptons
  produced in Z decays}'',} \textit{ Z. Phys. C} \textbf{ 74} (1997) 57,
  \href{http://dx.doi.org/10.1007/s002880050370}{\doi{10.1007/s002880050370}}.
[Erratum: {\it Z. Phys. C} {\bf 75} (1997) 580].

\bibitem{Vaitaitis:1999wq}
\hrefCMSnoop {}{{NuTeV (E815)} Collaboration, ``{Search for neutral heavy
  leptons in a high-energy neutrino beam}'',} \textit{ Phys. Rev. Lett.}
  \textbf{ 83} (1999) 4943,
  \href{http://dx.doi.org/10.1103/PhysRevLett.83.4943}{\doi{10.1103/PhysRevLett.83.4943}},
\href{http://www.arXiv.org/abs/hep-ex/9908011}{\texttt{arXiv:hep-ex/9908011}}.

\bibitem{Acciarri:1999qj}
\hrefCMSnoop {}{{L3} Collaboration, ``{Search for heavy isosinglet neutrinos in
  $\EE$ annihilation at $130 < \sqrt{s} < 189\GeV$}'',} \textit{ Phys. Lett. B}
  \textbf{ 461} (1999) 397,
  \href{http://dx.doi.org/10.1016/S0370-2693(99)00852-7}{\doi{10.1016/S0370-2693(99)00852-7}},
\href{http://www.arXiv.org/abs/hep-ex/9909006}{\texttt{arXiv:hep-ex/9909006}}.

\bibitem{Achard:2001qv}
\hrefCMSnoop {}{{L3} Collaboration, ``{Search for heavy isosinglet neutrino in
  $\EE$ annihilation at LEP}'',} \textit{ Phys. Lett. B} \textbf{ 517} (2001)
  67,
  \href{http://dx.doi.org/10.1016/S0370-2693(01)00993-5}{\doi{10.1016/S0370-2693(01)00993-5}},
\href{http://www.arXiv.org/abs/hep-ex/0107014}{\texttt{arXiv:hep-ex/0107014}}.

\bibitem{Liventsev:2013zz}
\hrefCMSnoop {}{{Belle} Collaboration, ``{Search for heavy neutrinos at
  Belle}'',} \textit{ Phys. Rev. D} \textbf{ 87} (2013) 071102,
  \href{http://dx.doi.org/10.1103/PhysRevD.87.071102}{\doi{10.1103/PhysRevD.87.071102}},
  \href{http://www.arXiv.org/abs/1301.1105}{\texttt{arXiv:1301.1105}}.
[Erratum: {\it Phys. Rev. D} {\bf 95} (2017) 099903,
  \DOI{10.1103/PhysRevD.95.099903}].

\bibitem{Aaij:2014aba}
\hrefCMSnoop {}{{LHCb Collaboration}, ``{Search for Majorana Neutrinos in ${\rm
  B}^- \to \pi^+\mu^-\mu^-$ Decays}'',} \textit{ Phys. Rev. Lett.} \textbf{
  112} (2014) 131802,
  \href{http://dx.doi.org/10.1103/PhysRevLett.112.131802}{\doi{10.1103/PhysRevLett.112.131802}},
\href{http://www.arXiv.org/abs/1401.5361}{\texttt{arXiv:1401.5361}}.

\bibitem{delAguila:2008pw}
\hrefCMSnoop {}{F.~del Aguila, J.~de~Blas, and M.~Perez-Victoria, ``{Effects of
  new leptons in electroweak precision data}'',} \textit{ Phys. Rev. D}
  \textbf{ 78} (2008) 013010,
  \href{http://dx.doi.org/10.1103/PhysRevD.78.013010}{\doi{10.1103/PhysRevD.78.013010}},
\href{http://www.arXiv.org/abs/0803.4008}{\texttt{arXiv:0803.4008}}.

\bibitem{Akhmedov:2013hec}
E.~Akhmedov\hrefCMSnoop {}{ {et~al.}, ``{Improving electro-weak fits with
  TeV-scale sterile neutrinos}'',} \textit{ JHEP} \textbf{ 05} (2013) 081,
  \href{http://dx.doi.org/10.1007/JHEP05(2013)081}{\doi{10.1007/JHEP05(2013)081}},
\href{http://www.arXiv.org/abs/1302.1872}{\texttt{arXiv:1302.1872}}.

\bibitem{deBlas:2013gla}
\hrefCMSnoop {}{J.~de~Blas, ``{Electroweak limits on physics beyond the
  Standard Model}'',} \textit{ Eur. Phys. J. Web Conf.} \textbf{ 60} (2013)
  19008,
  \href{http://dx.doi.org/10.1051/epjconf/20136019008}{\doi{10.1051/epjconf/20136019008}},
\href{http://www.arXiv.org/abs/1307.6173}{\texttt{arXiv:1307.6173}}.

\bibitem{Basso:2013jka}
\hrefCMSnoop {}{L.~Basso, O.~Fischer, and J.~J. van~der Bij, ``{Precision tests
  of unitarity in leptonic mixing}'',} \textit{ Eur. Phys. Lett.} \textbf{ 105}
  (2014) 11001,
  \href{http://dx.doi.org/10.1209/0295-5075/105/11001}{\doi{10.1209/0295-5075/105/11001}},
\href{http://www.arXiv.org/abs/1310.2057}{\texttt{arXiv:1310.2057}}.

\bibitem{Antusch:2015mia}
\hrefCMSnoop {}{S.~Antusch and O.~Fischer, ``{Testing sterile neutrino
  extensions of the Standard Model at future lepton colliders}'',} \textit{
  JHEP} \textbf{ 05} (2015) 053,
  \href{http://dx.doi.org/10.1007/JHEP05(2015)053}{\doi{10.1007/JHEP05(2015)053}},
\href{http://www.arXiv.org/abs/1502.05915}{\texttt{arXiv:1502.05915}}.

\bibitem{Deppisch:2015qwa}
\hrefCMSnoop {}{F.~F. Deppisch, P.~S. Bhupal~Dev, and A.~Pilaftsis,
  ``{Neutrinos and collider physics}'',} \textit{ New J. Phys.} \textbf{ 17}
  (2015) 075019,
  \href{http://dx.doi.org/10.1088/1367-2630/17/7/075019}{\doi{10.1088/1367-2630/17/7/075019}},
\href{http://www.arXiv.org/abs/1502.06541}{\texttt{arXiv:1502.06541}}.

\bibitem{delAguila:2008cj}
\hrefCMSnoop {}{F.~del Aguila and J.~A. Aguilar-Saavedra, ``{Distinguishing
  seesaw models at LHC with multi-lepton signals}'',} \textit{ Nucl. Phys. B}
  \textbf{ 813} (2009) 22,
  \href{http://dx.doi.org/10.1016/j.nuclphysb.2008.12.029}{\doi{10.1016/j.nuclphysb.2008.12.029}},
\href{http://www.arXiv.org/abs/0808.2468}{\texttt{arXiv:0808.2468}}.

\bibitem{Das:2012ze}
\hrefCMSnoop {}{A.~Das and N.~Okada, ``{Inverse seesaw neutrino signatures at
  the LHC and ILC}'',} \textit{ Phys. Rev. D} \textbf{ 88} (2013) 113001,
  \href{http://dx.doi.org/10.1103/PhysRevD.88.113001}{\doi{10.1103/PhysRevD.88.113001}},
\href{http://www.arXiv.org/abs/1207.3734}{\texttt{arXiv:1207.3734}}.

\bibitem{Das:2014jxa}
\hrefCMSnoop {}{A.~Das, P.~S. Bhupal~Dev, and N.~Okada, ``{Direct bounds on
  electroweak scale pseudo-Dirac neutrinos from $\sqrt s=8$ TeV LHC data}'',}
  \textit{ Phys. Lett. B} \textbf{ 735} (2014) 364,
  \href{http://dx.doi.org/10.1016/j.physletb.2014.06.058}{\doi{10.1016/j.physletb.2014.06.058}},
\href{http://www.arXiv.org/abs/1405.0177}{\texttt{arXiv:1405.0177}}.

\bibitem{Izaguirre:2015pga}
\hrefCMSnoop {}{E.~Izaguirre and B.~Shuve, ``{Multilepton and lepton jet probes
  of sub-weak-scale right-handed neutrinos}'',} \textit{ Phys. Rev. D} \textbf{
  91} (2015) 093010,
  \href{http://dx.doi.org/10.1103/PhysRevD.91.093010}{\doi{10.1103/PhysRevD.91.093010}},
\href{http://www.arXiv.org/abs/1504.02470}{\texttt{arXiv:1504.02470}}.

\bibitem{Dib:2015oka}
\hrefCMSnoop {}{C.~O. Dib and C.~S. Kim, ``{Discovering sterile neutrinos
  lighter than $M_\PW$ at the LHC}'',} \textit{ Phys. Rev. D} \textbf{ 92}
  (2015) 093009,
  \href{http://dx.doi.org/10.1103/PhysRevD.92.093009}{\doi{10.1103/PhysRevD.92.093009}},
\href{http://www.arXiv.org/abs/1509.05981}{\texttt{arXiv:1509.05981}}.

\bibitem{Das:2016hof}
\hrefCMSnoop {}{A.~Das, P.~Konar, and S.~Majhi, ``{Production of Heavy neutrino
  in next-to-leading order QCD at the LHC and beyond}'',} \textit{ JHEP}
  \textbf{ 06} (2016) 019,
  \href{http://dx.doi.org/10.1007/JHEP06(2016)019}{\doi{10.1007/JHEP06(2016)019}},
\href{http://www.arXiv.org/abs/1604.00608}{\texttt{arXiv:1604.00608}}.

\bibitem{Dib:2016wge}
\hrefCMSnoop {}{C.~O. Dib, C.~S. Kim, K.~Wang, and J.~Zhang, ``{Distinguishing
  Dirac/Majorana sterile neutrinos at the LHC}'',} \textit{ Phys. Rev. D}
  \textbf{ 94} (2016) 013005,
  \href{http://dx.doi.org/10.1103/PhysRevD.94.013005}{\doi{10.1103/PhysRevD.94.013005}},
\href{http://www.arXiv.org/abs/1605.01123}{\texttt{arXiv:1605.01123}}.

\bibitem{Dib:2017iva}
\hrefCMSnoop {}{C.~O. Dib, C.~S. Kim, and K.~Wang, ``{Signatures of Dirac and
  Majorana sterile neutrinos in trilepton events at the LHC}'',} \textit{ Phys.
  Rev. D} \textbf{ 95} (2017) 115020,
  \href{http://dx.doi.org/10.1103/PhysRevD.95.115020}{\doi{10.1103/PhysRevD.95.115020}},
\href{http://www.arXiv.org/abs/1703.01934}{\texttt{arXiv:1703.01934}}.

\bibitem{Dib:2017vux}
\hrefCMSnoop {}{C.~O. Dib, C.~S. Kim, and K.~Wang, ``{Search for heavy sterile
  neutrinos in trileptons at the LHC}'',} \textit{ Chin. Phys. C} \textbf{ 41}
  (2017) 103103,
  \href{http://dx.doi.org/10.1088/1674-1137/41/10/103103}{\doi{10.1088/1674-1137/41/10/103103}},
\href{http://www.arXiv.org/abs/1703.01936}{\texttt{arXiv:1703.01936}}.

\bibitem{Dube:2017jgo}
\hrefCMSnoop {}{S.~Dube, D.~Gadkari, and A.~M. Thalapillil, ``{Lepton jets and
  low-mass sterile neutrinos at hadron colliders}'',} \textit{ Phys. Rev. D}
  \textbf{ 96} (2017) 055031,
  \href{http://dx.doi.org/10.1103/PhysRevD.96.055031}{\doi{10.1103/PhysRevD.96.055031}},
\href{http://www.arXiv.org/abs/1707.00008}{\texttt{arXiv:1707.00008}}.

\bibitem{Das:2017gke}
\hrefCMSnoop {}{A.~Das, P.~Konar, and A.~Thalapillil, ``{Jet substructure
  shedding light on heavy Majorana neutrinos at the LHC}'',} \textit{ JHEP}
  \textbf{ 02} (2018) 083,
  \href{http://dx.doi.org/10.1007/JHEP02(2018)083}{\doi{10.1007/JHEP02(2018)083}},
\href{http://www.arXiv.org/abs/1709.09712}{\texttt{arXiv:1709.09712}}.

\bibitem{Arbelaez:2017zqq}
\hrefCMSnoop {}{C.~Arbela\'{e}z, C.~Dib, I.~Schmidt, and J.~C. Vasquez,
  ``{Probing the Dirac or Majorana nature of the heavy neutrinos in pure
  leptonic decays at the LHC}'',} \textit{ Phys. Rev. D} \textbf{ 97} (2018)
  055011,
  \href{http://dx.doi.org/10.1103/PhysRevD.97.055011}{\doi{10.1103/PhysRevD.97.055011}},
\href{http://www.arXiv.org/abs/1712.08704}{\texttt{arXiv:1712.08704}}.

\bibitem{Bhardwaj:2018lma}
\hrefCMSnoop {}{A.~Bhardwaj, A.~Das, P.~Konar, and A.~Thalapillil,
  ``{Challenging Sterile Neutrino Searches at the LHC Complemented by Jet
  Substructure Techniques}'',} (2018).
\href{http://www.arXiv.org/abs/1801.00797}{\texttt{arXiv:1801.00797}}.

\bibitem{Chatrchyan:2008zzk}
\hrefCMSnoop {}{{CMS Collaboration}, ``The {CMS} experiment at the {CERN}
  {LHC}'',} \textit{ JINST} \textbf{ 3} (2008) S08004,
  \href{http://dx.doi.org/10.1088/1748-0221/3/08/S08004}{\doi{10.1088/1748-0221/3/08/S08004}}.

\bibitem{Khachatryan:2016bia}
\hrefCMSnoop {}{{CMS Collaboration}, ``{The CMS trigger system}'',} \textit{
  JINST} \textbf{ 12} (2017) P01020,
  \href{http://dx.doi.org/10.1088/1748-0221/12/01/P01020}{\doi{10.1088/1748-0221/12/01/P01020}},
\href{http://www.arXiv.org/abs/1609.02366}{\texttt{arXiv:1609.02366}}.

\bibitem{MADGRAPH5}
J.~Alwall\hrefCMSnoop {}{ {et~al.}, ``{The automated computation of tree-level
  and next-to-leading order differential cross sections, and their matching to
  parton shower simulations}'',} \textit{ JHEP} \textbf{ 07} (2014) 079,
  \href{http://dx.doi.org/10.1007/JHEP07(2014)079}{\doi{10.1007/JHEP07(2014)079}},
\href{http://www.arXiv.org/abs/1405.0301}{\texttt{arXiv:1405.0301}}.

\bibitem{Campbell:2010ff}
\hrefCMSnoop {}{J.~M. Campbell and R.~K. Ellis, ``{MCFM for the Tevatron and
  the LHC}'',} \textit{ Nucl. Phys. Proc. Suppl.} \textbf{ 205} (2010) 10,
  \href{http://dx.doi.org/10.1016/j.nuclphysbps.2010.08.011}{\doi{10.1016/j.nuclphysbps.2010.08.011}},
\href{http://www.arXiv.org/abs/1007.3492}{\texttt{arXiv:1007.3492}}.

\bibitem{Melia:2011tj}
\hrefCMSnoop {}{T.~Melia, P.~Nason, R.~Rontsch, and G.~Zanderighi, ``{$\PW^+
  \PW^-$, $\PW \PZ$ and $\PZ \PZ$ production in the POWHEG BOX}'',} \textit{
  JHEP} \textbf{ 11} (2011) 078,
  \href{http://dx.doi.org/10.1007/JHEP11(2011)078}{\doi{10.1007/JHEP11(2011)078}},
\href{http://www.arXiv.org/abs/1107.5051}{\texttt{arXiv:1107.5051}}.

\bibitem{Nason:2013ydw}
\hrefCMSnoop {}{P.~Nason and G.~Zanderighi, ``{$\PW^+ \PW^-$, $\PW \PZ$ and
  $\PZ \PZ$ production in the POWHEG-BOX-V2}'',} \textit{ Eur. Phys. J. C}
  \textbf{ 74} (2014) 2702,
  \href{http://dx.doi.org/10.1140/epjc/s10052-013-2702-5}{\doi{10.1140/epjc/s10052-013-2702-5}},
\href{http://www.arXiv.org/abs/1311.1365}{\texttt{arXiv:1311.1365}}.

\bibitem{Ball:2014uwa}
\hrefCMSnoop {}{{NNPDF} Collaboration, ``{Parton distributions for the LHC Run
  II}'',} \textit{ JHEP} \textbf{ 04} (2015) 040,
  \href{http://dx.doi.org/10.1007/JHEP04(2015)040}{\doi{10.1007/JHEP04(2015)040}},
\href{http://www.arXiv.org/abs/1410.8849}{\texttt{arXiv:1410.8849}}.

\bibitem{Sjostrand:2007gs}
\hrefCMSnoop {}{T.~Sj{\"o}strand, S.~Mrenna, and P.~Z. Skands, ``{A brief
  introduction to PYTHIA 8.1}'',} \textit{ Comput. Phys. Commun.} \textbf{ 178}
  (2008) 852,
  \href{http://dx.doi.org/10.1016/j.cpc.2008.01.036}{\doi{10.1016/j.cpc.2008.01.036}},
\href{http://www.arXiv.org/abs/0710.3820}{\texttt{arXiv:0710.3820}}.

\bibitem{Skands:2014pea}
\hrefCMSnoop {}{P.~Skands, S.~Carrazza, and J.~Rojo, ``{Tuning PYTHIA 8.1: the
  Monash 2013 tune}'',} \textit{ Eur. Phys. J. C} \textbf{ 74} (2014) 3024,
  \href{http://dx.doi.org/10.1140/epjc/s10052-014-3024-y}{\doi{10.1140/epjc/s10052-014-3024-y}},
\href{http://www.arXiv.org/abs/1404.5630}{\texttt{arXiv:1404.5630}}.

\bibitem{CMS-PAS-GEN-14-001}
\hrefCMSnoop {}{{CMS Collaboration}, ``{Event generator tunes obtained from
  underlying event and multiparton scattering measurements}'',} \textit{ Eur.
  Phys. J. C} \textbf{ 76} (2016) 155,
  \href{http://dx.doi.org/10.1140/epjc/s10052-016-3988-x}{\doi{10.1140/epjc/s10052-016-3988-x}},
\href{http://www.arXiv.org/abs/1512.00815}{\texttt{arXiv:1512.00815}}.

\bibitem{Alwall:2007fs}
J.~Alwall\hrefCMSnoop {}{ {et~al.}, ``{Comparative study of various algorithms
  for the merging of parton showers and matrix elements in hadronic
  collisions}'',} \textit{ Eur. Phys. J. C} \textbf{ 53} (2008) 473,
  \href{http://dx.doi.org/10.1140/epjc/s10052-007-0490-5}{\doi{10.1140/epjc/s10052-007-0490-5}},
\href{http://www.arXiv.org/abs/0706.2569}{\texttt{arXiv:0706.2569}}.

\bibitem{Frederix:2012ps}
\hrefCMSnoop {}{R.~Frederix and S.~Frixione, ``{Merging meets matching in
  MC@NLO}'',} \textit{ JHEP} \textbf{ 12} (2012) 061,
  \href{http://dx.doi.org/10.1007/JHEP12(2012)061}{\doi{10.1007/JHEP12(2012)061}},
\href{http://www.arXiv.org/abs/1209.6215}{\texttt{arXiv:1209.6215}}.

\bibitem{Degrande:2016aje}
\hrefCMSnoop {}{C.~Degrande, O.~Mattelaer, R.~Ruiz, and J.~Turner, ``{Fully
  automated precision predictions for heavy neutrino production mechanisms at
  hadron colliders}'',} \textit{ Phys. Rev. D} \textbf{ 94} (2016) 053002,
  \href{http://dx.doi.org/10.1103/PhysRevD.94.053002}{\doi{10.1103/PhysRevD.94.053002}},
\href{http://www.arXiv.org/abs/1602.06957}{\texttt{arXiv:1602.06957}}.

\bibitem{Alva:2014gxa}
\hrefCMSnoop {}{D.~Alva, T.~Han, and R.~Ruiz, ``{Heavy Majorana neutrinos from
  $\PW\gamma$ fusion at hadron colliders}'',} \textit{ JHEP} \textbf{ 02}
  (2015) 072,
  \href{http://dx.doi.org/10.1007/JHEP02(2015)072}{\doi{10.1007/JHEP02(2015)072}},
\href{http://www.arXiv.org/abs/1411.7305}{\texttt{arXiv:1411.7305}}.

\bibitem{Manohar:2016nzj}
\hrefCMSnoop {}{A.~Manohar, P.~Nason, G.~P. Salam, and G.~Zanderighi, ``{How
  bright is the proton? A precise determination of the photon parton
  distribution function}'',} \textit{ Phys. Rev. Lett.} \textbf{ 117} (2016)
  242002,
  \href{http://dx.doi.org/10.1103/PhysRevLett.117.242002}{\doi{10.1103/PhysRevLett.117.242002}},
\href{http://www.arXiv.org/abs/1607.04266}{\texttt{arXiv:1607.04266}}.

\bibitem{Geant}
\hrefCMSnoop {}{{GEANT4} Collaboration, ``{GEANT4 --- a simulation toolkit}'',}
  \textit{ Nucl. Instrum. Meth. A} \textbf{ 506} (2003) 250,
\href{http://dx.doi.org/10.1016/S0168-9002(03)01368-8}{\doi{10.1016/S0168-9002(03)01368-8}}.

\bibitem{Sirunyan:2017ulk}
\hrefCMSnoop {}{{CMS Collaboration}, ``{Particle-flow reconstruction and global
  event description with the CMS detector}'',} \textit{ JINST} \textbf{ 12}
  (2017) P10003,
  \href{http://dx.doi.org/10.1088/1748-0221/12/10/P10003}{\doi{10.1088/1748-0221/12/10/P10003}},
\href{http://www.arXiv.org/abs/1706.04965}{\texttt{arXiv:1706.04965}}.

\bibitem{Cacciari:2008gp}
\hrefCMSnoop {}{M.~Cacciari, G.~P. Salam, and G.~Soyez, ``{The anti-$k_{t}$ jet
  clustering algorithm}'',} \textit{ JHEP} \textbf{ 04} (2008) 063,
  \href{http://dx.doi.org/10.1088/1126-6708/2008/04/063}{\doi{10.1088/1126-6708/2008/04/063}},
\href{http://www.arXiv.org/abs/0802.1189}{\texttt{arXiv:0802.1189}}.

\bibitem{Cacciari:fastjet1}
\hrefCMSnoop {}{M.~Cacciari, G.~P. Salam, and G.~Soyez, ``{FastJet user
  manual}'',} \textit{ Eur. Phys. J. C} \textbf{ 72} (2012) 1896,
  \href{http://dx.doi.org/10.1140/epjc/s10052-012-1896-2}{\doi{10.1140/epjc/s10052-012-1896-2}},
\href{http://www.arXiv.org/abs/1111.6097}{\texttt{arXiv:1111.6097}}.

\bibitem{Cacciari:fastjet2}
\hrefCMSnoop {}{M.~Cacciari and G.~P. Salam, ``{Dispelling the $N^{3}$ myth for
  the $k_{t}$ jet-finder}'',} \textit{ Phys. Lett. B} \textbf{ 641} (2006) 57,
  \href{http://dx.doi.org/10.1016/j.physletb.2006.08.037}{\doi{10.1016/j.physletb.2006.08.037}},
\href{http://www.arXiv.org/abs/hep-ph/0512210}{\texttt{arXiv:hep-ph/0512210}}.

\bibitem{Chatrchyan:2011ds}
\hrefCMSnoop {}{{CMS Collaboration}, ``{Determination of jet energy calibration
  and transverse momentum resolution in CMS}'',} \textit{ JINST} \textbf{ 6}
  (2011) P11002,
  \href{http://dx.doi.org/10.1088/1748-0221/6/11/P11002}{\doi{10.1088/1748-0221/6/11/P11002}},
\href{http://www.arXiv.org/abs/1107.4277}{\texttt{arXiv:1107.4277}}.

\bibitem{Khachatryan:2016kdb}
\hrefCMSnoop {}{{CMS Collaboration}, ``{Jet energy scale and resolution in the
  CMS experiment in pp collisions at 8 TeV}'',} \textit{ JINST} \textbf{ 12}
  (2017) P02014,
  \href{http://dx.doi.org/10.1088/1748-0221/12/02/P02014}{\doi{10.1088/1748-0221/12/02/P02014}},
\href{http://www.arXiv.org/abs/1607.03663}{\texttt{arXiv:1607.03663}}.

\bibitem{CMS-PAS-JME-16-003}
\href {https://cds.cern.ch/record/2256875}{{CMS Collaboration}, ``{Jet
  algorithms performance in 13 TeV data}'',} CMS Physics Analysis Summary
  CMS-PAS-JME-16-003, 2010.

\bibitem{Chatrchyan:2012jua}
\hrefCMSnoop {}{{CMS Collaboration}, ``{Identification of b-quark jets with the
  CMS experiment}'',} \textit{ JINST} \textbf{ 8} (2013) P04013,
  \href{http://dx.doi.org/10.1088/1748-0221/8/04/P04013}{\doi{10.1088/1748-0221/8/04/P04013}},
\href{http://www.arXiv.org/abs/1211.4462}{\texttt{arXiv:1211.4462}}.

\bibitem{BTV-16-002}
\hrefCMSnoop {}{{CMS Collaboration}, ``{Identification of heavy-flavour jets
  with the CMS detector in pp collisions at 13 TeV}'',} (2017).
  \href{http://www.arXiv.org/abs/1712.07158}{\texttt{arXiv:1712.07158}}.
Submitted to {\it JINST}.

\bibitem{CMS-PAS-JME-16-004}
\href {https://cds.cern.ch/record/2205284}{{CMS Collaboration}, ``{Performance
  of missing energy reconstruction in 13 TeV pp collision data using the CMS
  detector}'',} CMS Physics Analysis Summary CMS-PAS-JME-16-004, 2016.

\bibitem{Khachatryan:2015hwa}
\hrefCMSnoop {}{{CMS Collaboration}, ``{Performance of electron reconstruction
  and selection with the CMS detector in proton-proton collisions at $\sqrt{s}
  = 8\TeV$}'',} \textit{ JINST} \textbf{ 10} (2015) P06005,
  \href{http://dx.doi.org/10.1088/1748-0221/10/06/P06005}{\doi{10.1088/1748-0221/10/06/P06005}},
\href{http://www.arXiv.org/abs/1502.02701}{\texttt{arXiv:1502.02701}}.

\bibitem{Chatrchyan:2012xi}
\hrefCMSnoop {}{{CMS Collaboration}, ``{Performance of CMS muon reconstruction
  in pp collision events at $\sqrt{s} = 7\TeV$}'',} \textit{ JINST} \textbf{ 7}
  (2012) P10002,
  \href{http://dx.doi.org/10.1088/1748-0221/7/10/P10002}{\doi{10.1088/1748-0221/7/10/P10002}},
\href{http://www.arXiv.org/abs/1206.4071}{\texttt{arXiv:1206.4071}}.

\bibitem{SUS-15-008}
\hrefCMSnoop {}{{CMS Collaboration}, ``{Search for new physics in same-sign
  dilepton events in proton-proton collisions at $\sqrt{s} = 13\TeV$}'',}
  \textit{ Eur. Phys. J. C} \textbf{ 76} (2016) 439,
  \href{http://dx.doi.org/10.1140/epjc/s10052-016-4261-z}{\doi{10.1140/epjc/s10052-016-4261-z}},
\href{http://www.arXiv.org/abs/1605.03171}{\texttt{arXiv:1605.03171}}.

\bibitem{Sirunyan:2017uyt}
\hrefCMSnoop {}{{CMS Collaboration}, ``{Search for physics beyond the standard
  model in events with two leptons of same sign, missing transverse momentum,
  and jets in proton-proton collisions at $\sqrt{s}=13\TeV$}'',} \textit{ Eur.
  Phys. J. C} \textbf{ 77} (2017) 578,
  \href{http://dx.doi.org/10.1140/epjc/s10052-017-5079-z}{\doi{10.1140/epjc/s10052-017-5079-z}},
\href{http://www.arXiv.org/abs/1704.07323}{\texttt{arXiv:1704.07323}}.

\bibitem{Sirunyan:2017vio}
\hrefCMSnoop {}{{CMS Collaboration}, ``{Measurement of the underlying event
  activity in inclusive Z boson production in proton-proton collisions at
  $\sqrt{s} = $ 13 TeV}'',} (2017).
  \href{http://www.arXiv.org/abs/1711.04299}{\texttt{arXiv:1711.04299}}.
{Submitted to {\it JHEP}}.

\bibitem{CMS-PAS-LUM-17-001}
\href {https://cds.cern.ch/record/2138682}{{CMS Collaboration}, ``{CMS
  Luminosity Measurement for the 2016 Data Taking Period}'',} CMS Physics
  Analysis Summary CMS-PAS-LUM-17-001, 2017.

\bibitem{Butterworth:2015oua}
\hrefCMSnoop {}{J.~Butterworth {et~al.}, ``{PDF4LHC recommendations for LHC Run
  II}'',} \textit{ J. Phys. G} \textbf{ 43} (2016) 023001,
  \href{http://dx.doi.org/10.1088/0954-3899/43/2/023001}{\doi{10.1088/0954-3899/43/2/023001}},
\href{http://www.arXiv.org/abs/1510.03865}{\texttt{arXiv:1510.03865}}.

\bibitem{Junk:1999kv}
\hrefCMSnoop {}{T.~Junk, ``{Confidence level computation for combining searches
  with small statistics}'',} \textit{ Nucl. Instrum. Meth. A} \textbf{ 434}
  (1999) 435,
  \href{http://dx.doi.org/10.1016/S0168-9002(99)00498-2}{\doi{10.1016/S0168-9002(99)00498-2}},
\href{http://www.arXiv.org/abs/hep-ex/9902006}{\texttt{arXiv:hep-ex/9902006}}.

\bibitem{Read:2002hq}
\hrefCMSnoop {}{A.~L. Read, ``{Presentation of search results: The
  CL$_\text{s}$ technique}'',} \textit{ J. Phys. G} \textbf{ 28} (2002) 2693,
\href{http://dx.doi.org/10.1088/0954-3899/28/10/313}{\doi{10.1088/0954-3899/28/10/313}}.

\bibitem{Cowan:2010js}
\hrefCMSnoop {}{G.~Cowan, K.~Cranmer, E.~Gross, and O.~Vitells, ``{Asymptotic
  formulae for likelihood-based tests of new physics}'',} \textit{ Eur. Phys.
  J. C} \textbf{ 71} (2011) 1554,
  \href{http://dx.doi.org/10.1140/epjc/s10052-011-1554-0}{\doi{10.1140/epjc/s10052-011-1554-0}},
  \href{http://www.arXiv.org/abs/1007.1727}{\texttt{arXiv:1007.1727}}.
[Erratum: {\it Eur. Phys. J. C} {\bf 73} (2013) 2501,
  \DOI{10.1140/epjc/s10052-013-2501-z}].

\bibitem{ATLAS:1379837}
\href {http://cdsweb.cern.ch/record/1379837}{{ATLAS and CMS Collaborations},
  ``Procedure for the {LHC} higgs boson search combination in summer 2011'',}
  Technical Report ATL-PHYS-PUB-2011-11, CMS-NOTE-2011-005, 2011.

\end{thebibliography}\endgroup
